\documentclass[%
 reprint,
 amsmath,amssymb,
 aps,
]{revtex4-2}

\usepackage{graphicx}
\usepackage{dcolumn}
\usepackage{bm}
\usepackage{hyperref}
\usepackage[table,xcdraw]{xcolor}
\usepackage{booktabs}
\usepackage{multirow}
\usepackage[normalem]{ulem}
\usepackage{subcaption}
\usepackage{caption}
\usepackage{ragged2e} 
\usepackage{comment}
\usepackage{tabularray}
\usepackage{makecell}
\usepackage{color, soul}

\newcommand{\ONR}[1]{$\mathcal{O}^{\rm NR}_{#1}$}
\font\bb=bbmss10 scaled 1200

\begin{document}

\preprint{APS/123-QED}

\title{Impact of nuclear shell model uncertainties on \\
silicon and germanium WIMP-nucleus limits
}%

\author{M. J. Zurowski}
 \email{madeleine.zurowski@utoronto.ca}
\affiliation{%
Department of Physics, University of Toronto, Toronto, ON M5S 1A7, Canada
}%

\author{R. Abdel Khaleq}
 \email{raghda.abdelkhaleq@anu.edu.au}
\affiliation{
 Department of Fundamental and Theoretical Physics,\\
Research School of Physics, Australian National University, ACT, 2601, Australia}%
\affiliation{
Department of Nuclear Physics and Accelerator Applications, Research School of Physics, Australian National University, ACT, 2601, Australia
}%
\affiliation{%
ARC Centre of Excellence for Dark Matter Particle Physics, Australia
}%

\date{\today}

\begin{abstract}
Dark matter (DM) direct detection searches look for nuclear recoils as a result of weakly interacting massive particles (WIMPs) scattering off nuclei. Calculating the rate associated with this WIMP-nucleus elastic scattering process involves input from several components, which include the direct detection experimental response functions, the high energy physics content employed, the DM halo velocity distribution, and the target nuclear structure information. Uncertainties in any of these components can impact the interpretation of experimental signals, and affect analysis of rate predictions. In this work, we focus on the uncertainties present due to the nuclear structure modelling of silicon and germanium targets, where a comprehensive set of nuclear form factors are employed through a non-relativistic effective field theory (NREFT) formalism. We show that these uncertainties, obtained from large-scale shell model calculations, impact the nuclear form factors and thus experimental exclusion limits for a SuperCDMS-like experiment. The impact of high energy physics content (through particle physics coefficients) on the magnitude of the nuclear uncertainties is also explored for several WIMP-nucleus scattering responses. Non-negligible nuclear uncertainties are present for several direct detection nuclear responses, indicating that nuclear structure must be accounted for in experimental analysis and interpretation.
\end{abstract}
\maketitle

\section{Introduction}

The nature of dark matter (DM) continues to elude our understanding, with the DM problem inspiring many decades of study. Among the many candidates proposed to describe the nature of DM is the Weakly Interacting Massive Particle (WIMP), a particle candidate with a mass in the GeV -- TeV range able to interact non-relativistically with nuclear targets in dark matter direct detection experiments. In this case, the DM-nucleus elastic scattering rate depends on several factors: the DM halo velocity distribution; the high energy physics interaction model; response functions describing the direct detection experiment of interest; and the target nuclear structure information. Efforts aimed at improving the modelling of each of these components contribute to reduced uncertainties in DM-nucleus scattering predictions important for experimental analysis and interpretation. 

We focus here on the nuclear structure modelling uncertainties, which may have a significant impact on the evaluation of rates and experimental exclusion limits given the assumption of different particle interaction models. The standard approach for modelling the WIMP nuclear operators involves considering both a spin-independent (SI) and spin-dependent (SD) response. This was expanded on by Fitzpatrick \textit{et al}. \cite{Fitzpatrick:2012ix,Anand:2013yka} through a non-relativistic effective field theory (NREFT) framework, to account for the orbital momentum of nucleons within the nucleus. As a result, additional nuclear responses are considered -- one depends on the orbital angular momentum $L$ and is hence $L$-dependent (LD), and another is both $s$ (spin)- and $L$-dependent (LSD). This more complete set of non-relativistic (NR) nuclear operators beyond the standard SD and SI interactions requires a more comprehensive treatment of the nuclear structure of targets, where the characterisation of the nuclear uncertainties present becomes important. This nuclear uncertainty characterisation can be quantified using large-scale shell model calculations for silicon and germanium, where various shell model interactions are employed to understand the impact of differences arising from nuclear modelling. The nuclear form factor uncertainties are compared with those in the exclusion limits for each nuclear operator, and the degree of uncertainty propagation present is discussed.   

This paper aims to provide a prescription to assess the impact of uncertainties in nuclear structure on experimental limits. This is performed specifically for silicon and germanium targets under the assumption of SuperCDMS-like experimental parameters.
This paper proceeds by providing a review of how a DM interaction rate is calculated in Sec. \ref{sec:rate_construction}, with Sec. \ref{sec:det_effects} discussing the transformation of this rate to the counterpart observed by a detector. Section \ref{sec:nuclear_struc} gives an overview of the available nuclear structure models of interest for the targets explored in this paper. In the relevant sections, we will also discuss the typical assumptions made for each contribution as well as the associated uncertainties or possible deviations from the standard. The impact of these on the final observable is presented in Sec. \ref{sec:results} and our conclusions in Sec. \ref{sec:conc}.

\section{Constructing the dark matter interaction rate}
\label{sec:rate_construction}

DM-nucleus elastic scattering has kinematics of the form $\chi + T \rightarrow \chi + T$, where the DM particle is denoted by $\chi$ and the target nucleus by $T$. This interaction is decomposed into a sum over all possible interactions between the WIMP and the nucleons $N$ within the nucleus. The momentum transfer in this process is given by \cite{Fitzpatrick:2012ix,Anand:2013yka}

\begin{equation}\label{momentum transfer}
    \vec{q}= \vec{p}\hspace{0.5mm}'-\vec{p}= \vec{k}-\vec{k}',
\end{equation}

\noindent where $\vec{p}\hspace{0.5mm}'$ ($\vec{p}$) is the outgoing (incoming) $\chi$ momentum and $\vec{k}'$ ($\vec{k}$) is the outgoing (incoming) $N$ momentum. The dark matter interaction rate then describes the number of DM interactions that could occur in a given target material, and is given by 
\begin{equation}
    \frac{{\rm d}R}{{\rm d}E_R} = N_T\frac{\rho}{m_{\chi}} \int_{v_{\rm min}}^{v_{\rm esc}} v f_{\rm lab}(\vec{v})\frac{{\rm d}\sigma_T}{{\rm d}E_R}{\rm d}^3v,
\label{eq:gen_rate}    
\end{equation}

\noindent where the minimum DM velocity required to produce a nuclear recoil energy $E_R$ in an elastic scattering process has the form

\begin{equation}
    v_{\text{min}} (E_R) = \frac{q}{2\mu_T} = \frac{1}{\mu_T} \sqrt{\frac{m_T E_R}{2}}.
\end{equation}

Here, $\mu_T= m_T m_\chi/(m_T+m_\chi)$ is the DM-nucleus reduced mass. The scattering cross-section $\sigma_T$ includes information on the nuclear structure and the high energy physics content employed (through the quark-level coupling coefficients), and is encoded with the physics which dictates the nature of the DM interaction. 

The usual approach for calculating the differential cross section is to use non-relativistic effective field theory (NREFT) operators to compute a scattering matrix element. This then gives the differential cross section
\begin{equation}
\label{eq:diff_xsec}
\begin{split}
    \frac{{\rm d}\sigma_T}{{\rm d}E_R} 
        &=A_{\rm prop}
            \left[\frac{1}{2j_{\chi}+1}\frac{1}{2j_{T}+1}     
            \sum_{\text{spins}}|\mathcal{M}|^2\right]\\
        &=A_{\rm prop}
            \sum_{i,j} \sum_{N,N'=p,n}  c^N_{i,{\rm NR}} c^{N'}_{j,{\rm NR}} F_{ij}^{(N,N')} (v^2,q^2),
\end{split}
\end{equation}

\noindent where $N, N'=\{p, n\}$ represents the proton ($p$) and neutron ($n$) components of the responses.
The term of proportionality $A_{\rm prop}$ will depend on the units one assumes for both the coupling coefficients $c^N_{i,{\rm NR}}$ and form factors $F_{ij}^{(N,N')} (v^2,q^2)$. For the remainder of this paper, we will use a `unitless' construction where the proportionality term is given by
\begin{equation}
    A_{\rm prop} = \frac{m_T}{2\pi v^2}.
\end{equation}
The full derivations of this are detailed in Refs.~\cite{Fitzpatrick:2012ix,Anand:2013yka,Cirelli_2013} where we direct the reader for details. In Table \ref{tab:nr_operators} we present the expressions for the `unitless' non-relativistic operators $\mathcal{O}_i^{\rm NR}$ (used in Ref.~\cite{Anand:2013yka}). In Appendix \ref{sec:ff_units} we include a detailed discussion that compares the `unitless' with the `unitful' operators (used in Refs.~\cite{Fitzpatrick:2012ix,Cirelli_2013}) and provide a method to convert between the two.

\begin{table}[!h]
\captionsetup{justification=Justified}
\caption{The non-relativistic operators $\mathcal{O}_i^{\rm NR}$ assumed in the unitless \cite{Anand:2013yka} expressions. The relative velocity is given by $\vec{v}^{\perp} \equiv \vec{v} + \vec{q}/2 \mu_N$, where $\mu_N= m_N m_\chi/(m_N+m_\chi)$ and $m_N$ is the nucleon mass.
}
\label{tab:nr_operators}
\begin{tabular}{@{}cl@{}}
\toprule
\textbf{Operator} & \textbf{Unitless expression} \\ \midrule
\ONR{1} & \bb 1 \\
\ONR{3} & $i \, \vec{s}_N \cdot \left(\frac{\vec{q}}{m_N} \times \vec{v}^\perp \right)$ \\
\ONR{4} & $\vec{s}_\chi \cdot \vec{s}_N$ \\
\ONR{5} &  $i \, \vec{s}_\chi \cdot \left(\frac{\vec{q}}{m_N} \times \vec{v}^\perp \right)$ \\
\ONR{6} &  $\left(\vec{s}_\chi \cdot \frac{\vec{q}}{m_N}\right) \left(\vec{s}_N \cdot \frac{\vec{q}}{m_N}\right)$ \\
\ONR{7} & $\vec{s}_N \cdot \vec{v}^\perp$ \\
\ONR{8} & $\vec{s}_\chi \cdot \vec{v}^\perp$ \\
\ONR{9} & $i \, \vec{s}_\chi \cdot \left(\vec{s}_N \times \frac{\vec{q}}{m_N}\right)$ \\
\ONR{10} & $i \, \vec{s}_N \cdot \frac{\vec{q}}{m_N}$ \\
\ONR{11} & $i \, \vec{s}_\chi \cdot \frac{\vec{q}}{m_N}$ \\
\ONR{12} & $\vec{v}^\perp \cdot \left(\vec{s}_\chi \times \vec{s}_N \right)$ \\ \bottomrule
\end{tabular}
\end{table}

These operators are used to calculate the form factors needed in Eq. \ref{eq:diff_xsec}. These form factors are given in Table \ref{tab:form_fact}, and depend on linear combinations of the nuclear responses $F^{(N,N')}_{X,Y} (q^2)$, where $X,Y \equiv$ \{$M$, $\Sigma''$, $\Sigma'$, $\Delta$, $\Phi''$, $\tilde{\Phi}'$\}.

\begin{table*}[t]
\captionsetup{justification=Justified}
\caption{Form factor expression for unitless non-relativistic operators. Note that $C(j_{\chi})=4j_{\chi}(j_{\chi}+1)/3$. For legibility, the $(N,N')$ superscript has been removed, but should be assumed for all $F_{X,Y}$ terms. We have also employed ${v^\perp_T}^2=v^2-v^2_{\rm min}$.
}
\label{tab:form_fact}
\begin{tabular}{@{}ll@{}}
\toprule
\textbf{Operator} &\textbf{Unitless expression} \\ \midrule
${F_{1,1}}$ & $F_{M}$ \\
${F_{3,3}}$ & $\frac{q^2}{8m_N^2} \left(v^2-v_{\rm min}^2\right)F_{\Sigma'} +\frac{q^4 }{4 m_N^4} F_{\Phi''}$ \\
${F_{4,4}}$ & $C(j_{\chi})\frac{1}{16} \left(F_{\Sigma''}+ F_{\Sigma'} \right)$ \\
${F_{5,5}}$ & $C(j_{\chi})\frac{1}{4} \left(\frac{q^4}{m_N^4} F_{\Delta} + \frac{q^2}{m_N^2} \left(v^2-v_{\rm min}^2\right)F_{M} \right)$ \\
${F_{6,6}}$ & $C(j_{\chi})\frac{q^4}{16m_N^4} F_{\Sigma''}$ \\
${F_{7,7}}$ & $\frac{1}{8} \left(v^2-v_{\rm min}^2\right)  F_{\Sigma'}$ \\
${F_{8,8}}$ & $C(j_{\chi})\frac{1}{4} \left(\frac{q^2}{m_N^2} F_{\Delta}+\left(v^2-v_{\rm min}^2\right) F_{M}\right)$ \\
${F_{9,9}}$ & $C(j_{\chi})\frac{q^2}{16m_N^2} F_{\Sigma'}$ \\
${F_{10,10}}$ & $\frac{q^2}{4m_N^2} F_{\Sigma''}$ \\
${F_{11,11}}$ & $C(j_{\chi})\frac{q^2}{4m_N^2} F_{M}$ \\
${F_{12,12}}$ & $C(j_{\chi}) \frac{1}{16}\left((v^2-v_{\rm min}^2)\left(\frac{1}{2}F_{\Sigma'}+F_{\Sigma''}\right)+\frac{q^2}{m_N^2}\left(F_{\tilde \Phi'}+F_{\Phi''}\right)\right)$ \\
${F_{1,3}}$ & $\frac{q^2}{2m^2_N}F_{M,\Phi''}$ \\
${F_{4,5}}$ & $C(j_{\chi})\frac{q^2}{8m^2_N}F_{\Sigma',\Delta}$ \\
${F_{4,6}}$ & $C(j_{\chi})\frac{q^2}{16m_N^2}F_{\Sigma''}$ \\
${F_{9,8}}$ & $-C(j_{\chi})\frac{q^2}{8m^2_N}F_{\Sigma',\Delta}$ \\
${F_{11,12}}$ & $C(j_{\chi})\frac{q^2}{8 m_N^2} F_{M, \Phi''}$ \\ \bottomrule
\end{tabular}
\end{table*}

With this general expression defined, we now lay out the assumptions used to select and calculate the coupling coefficients (Sec. \ref{sec:coupling_coeffs}), nuclear structure terms (Sec. \ref{sec:nuclear_resp}), and DM velocity distribution (Sec. \ref{sec:vel_dist}).

\subsection{Coupling coefficients and cross sections}
\label{sec:coupling_coeffs}
Rather than setting limits on individual coupling constants, experiments tend to set limits on an effective cross section. Thus, we want to be able to define a cross section associated with some set of coupling constants. For some models, this is a trivial mapping (particularly where only one operator is used), and examples of these are given in Refs.~\cite{Anand:2013yka, deap}. However, when multiple operators are assumed, there is no longer a simple, physical cross section. \\
One way of parametrising this when using the unitless expressions, where the coupling constants all have the same units, is to think of the combination of coupling constants as a vector that we can express with some overall normalisation and set of unit vectors \cite{Kang_2019, Zurowski_2020};
\begin{equation}
\begin{split}
    \vec{c_0} &= \{c_{i,{\rm NR}}^{N}\} \\
            &= c_0 \{\hat{c}_{i,{\rm NR}}^{N}\},\\
    \hat{c}^N_{i,{\rm NR}} &= \frac{c^N_{i,{\rm NR}} }{c_0}.
\end{split}
\label{coupling_vector}
\end{equation}
In this construction, the normalisation $c_0$ is a measure of the overall `strength' of the total interaction, while the `direction' of the vector gives information on the relative contribution of different types of couplings.
This allows us to express the differential cross section as
\begin{equation}
    \frac{{\rm d}\sigma_T}{{\rm d}E_R} 
        = \frac{m_T}{2\pi v^2} 
    c_0^2\sum_{i,j}\sum_{N, N'=n,p} \hat{c}_{i,{\rm NR}}^{N}\hat{c}_{j,{\rm NR}}^{N'}F^{(N,N')}_{ij}(v,q). 
\end{equation}
Adopting the parametrisation 
\begin{equation}
    \sigma \equiv \frac{\mu^2_p}{\pi}c_0^2,
\end{equation}
then allows the inclusion of a DM cross section $\sigma_{\chi}$:
\begin{equation}
    \frac{{\rm d}\sigma_T}{{\rm d}E_R} 
        = \frac{m_T \sigma_{\chi}}{2\mu^2_p v^2} \sum_{i,j}\sum_{N, N'=n,p} \hat{c}_{i,{\rm NR}}^{N}\hat{c}_{j,{\rm NR}}^{N'}F^{(N,N')}_{ij}(v,q).
\label{full_cross}    
\end{equation}
The subtle, but important, point here is we no longer depend on the absolute value of a coupling constant but instead on its value relative to some choice of normalisation $c_0$. The choice of this normalisation factor dictates how we should interpret the effective cross section $\sigma_{\chi}$. Typically, one of the following choices is made:

\begin{itemize}
    \item $\sigma_{\chi,N}$ -- the scattering cross section for DM off an atomic nucleus. Coupling constants are normalised to the {\it sum of all couplings} as if they formed a single `vector', so $c_0 = \sqrt{\sum_{i,N} \left(c_{i,{\rm NR}}^N\right)^2}$.
    \item $\sigma_{\chi,p}$ -- the scattering cross section for DM off a proton. Coupling constants are normalised to the {\it sum of proton couplings}, so $c_0 = \sqrt{\sum_{i} \left(c_{i,{\rm NR}}^p\right)^2}$.
    \item $\sigma_{\chi,n}$ -- the scattering cross section for DM off a neutron. Coupling constants are normalised to the {\it sum of neutron couplings}, so $c_0 = \sqrt{\sum_{i} \left(c_{i,{\rm NR}}^n\right)^2}$.
\end{itemize}

As an example, consider the traditional isoscalar spin-independent DM interaction model, where $c_1^p=c_1^n$ and all other couplings are zero. Limits for this are typically expressed on the proton cross section $\sigma_{\chi,p}$, which can be computed by taking $\hat{c}_1^p=\hat{c}_1^n=1$. If we wanted to instead compute limits on $\sigma_{\chi,N}$, we would take $\hat{c}_1^p=\hat{c}_1^n=1/\sqrt{2}$.

Similarly, for the isovector coupling where $c_1^p=-c_1^n$, one would have the following scenarios to choose from:
\begin{itemize}
    \item For limits on $\sigma_{\chi,N}$, set $\hat{c}_1^p=1/\sqrt{2}$ and $\hat{c}_1^n=-1/\sqrt{2}$.
    \item For limits on $\sigma_{\chi,p}$, set $\hat{c}_1^p=1$ and $\hat{c}_1^n=-1$.
\end{itemize}

The benefit of this method for expressing the coupling constants and cross section is it suits itself equally well to models that are constructed based on experimental constraints (such as xenophobic DM \cite{PhysRevD.88.015021,PhysRevD.89.016017}) and those that are derived from some higher energy model or set of quark couplings (see Ref.~\cite{Cirelli_2013} and Appendix \ref{sec:quark_couplings}) that may probe more than one NR operator simultaneously. 

\subsection{Nuclear form factors}\label{sec:nuclear_resp}

The nuclear form factors $F^{(N,N')}_{X,Y} (q^2)$ in Table~\ref{tab:form_fact} are associated with six nuclear responses, where $X,Y \equiv$ \{$M$, $\Sigma''$, $\Sigma'$, $\Delta$, $\Phi''$, $\tilde{\Phi}'$\}. Each response $X$ corresponds to a different aspect of nuclear structure, and thereby a different mechanism for WIMP-nucleus scattering. The explicit definitions of these responses can be found in Refs.~\cite{Fitzpatrick:2012ix,Anand:2013yka}, where the long-wavelength limit ($q \rightarrow 0$) of these responses provides an indication of their behaviour -- this is summarised in Table~\ref{tab:nuclear operator summary}.  

\begingroup
\renewcommand{\arraystretch}{1.25}
\begin{table}[htbp] 
\centering
\captionsetup{justification=Justified}
\caption{Summary of the long-wavelength limit behaviour of the nuclear operators $M$, $\Sigma''$, $\Sigma'$, $\Delta$, $\Phi''$, and $\tilde{\Phi}'$.
SI and SD are the standard spin-independent and spin-dependent operators, respectively. ``SO" refers to spin-orbit, LD represents angular momentum $L$ dependence, whilst LSD denotes dependence on both spin $s$ and $L$. Specifically, the SD operators $\Sigma''$, $\Sigma'$ correspond to longitudinal and transverse responses, respectively; $\Delta$ is associated with the orbital angular momentum; $\Phi''$ is associated with spin-orbit structure; and $\tilde{\Phi}'$ is a tensor term. $M$ is the standard SI response.
\label{tab:nuclear operator summary}}
\begin{tabular}{|c||c|c|c|c|c|c|}
\hline 
\textbf{Response} & $M$ & $\Sigma''$ & $\Sigma'$ & $\Delta$ & $\Phi''$ & $\tilde{\Phi}'$  \\
\hline 
\textbf{Behaviour} & SI & SD & SD & LD & LSD (SO) & LSD (tensor)  \\
\hline
\end{tabular}
\end{table}
\endgroup

The nuclear form factor $F^{(N,N')}_{X,Y} (q^2)$ takes on the expression \cite{Fitzpatrick:2012ix,Anand:2013yka}

\begin{equation}\label{FF eqn}
\resizebox{0.887\linewidth}{!}{%
    $F^{(N,N')}_{X,Y} (q^2)  \equiv \frac{4\pi}{2J_i+1} \sum_{J=0}^{2J_i} \langle J_i || X_J^{(N)} || J_i \rangle \langle J_i || Y^{(N')}_J || J_i \rangle,$}
\end{equation}

\noindent where $J_i$ is the nuclear ground state angular momentum with $|J_i \rangle$ being the wave function that represents this state. $N, N'=\{p, n\}$ represents the proton ($p$) and neutron ($n$) components of the nuclear response, where these operators are defined as 
\begin{equation}
\begin{split}
    X_J^{(p)} &= \frac{1+\tau_3}{2} \ X_J, \\
    X_J^{(n)} &= \frac{1-\tau_3}{2} \ X_J.
\end{split}    
\end{equation}

Here, $\tau_3$ is the nucleon isospin operator, where we have taken the isospin projections to be $m_t=+1/2$ for protons and $m_t=-1/2$ for neutrons for all calculations.

For the sake of notation, we define $F^{(N,N')}_X (q^2)  \equiv  F^{(N,N')}_{X,X} (q^2)$, and note that for these non-interference terms (where $X=Y$) the order of the nucleons does not matter, so $F^{(p,n)}_{X} (q^2)= F^{(n,p)}_{X} (q^2)$. These nuclear form factors hold all of the nuclear structure information about the nuclear targets of interest in an experiment, and can be used to gauge the impact of nuclear structure modelling uncertainties on WIMP-nucleus elastic scattering predictions and experimental interpretations. The cross terms (where $X\neq Y$) in Eq.~\ref{FF eqn} only exist for the nuclear response combinations $\{M_{JM}, \ \Phi''_{JM} \}$ and $\{\Sigma'_{JM}, \ \Delta_{JM}\}$.

In this work we consider a spherical shell model approach, where a spherical harmonic oscillator single-particle basis is employed. In this case, the nuclear response functions take on the form of $e^{-2y}p(y)$, with $p(y)$ being a polynomial where $y=(qb/2)^2$. Here, the harmonic oscillator length parameter is given by $b= {1}/{\sqrt{m_N \omega}} \approx \sqrt{41.467/(45A^{-1/3} - 25A^{-2/3})}$~fm, where $\omega$ is the oscillator frequency.

Evaluating these form factors involves writing the nuclear matrix elements in Eq.~\ref{FF eqn} as a product of single-nucleon matrix elements (which have analytic expressions) and One-Body Density Matrix Elements (OBDMEs)  $\Psi^{J;\tau}_{|\alpha|, |\beta|}$ (see Refs.~\cite{Fitzpatrick:2012ix,Anand:2013yka} for further details). These OBDMEs then hold all of the relevant nuclear structure information, and will take on different values for various nuclear targets. The nuclear uncertainty analysis is performed by evaluating these OBDMEs using various nuclear shell model interactions for a specific target, and comparing the resulting nuclear form factors obtained. Additional information on this process is provided in Sec.~\ref{sec:nuclear_struc}, and in Refs.~\cite{AbdelKhaleq:2023ipt,AbdelKhaleq:2024hir}.

We note that Table~\ref{tab:form_fact} only provides the expression for one of the cross terms $F_{i,j}^{(N,N')}$ for a specific set of $\{i,j\}$ values. To obtain the expression for the $\{j,i\}$ form factors, the $X,Y$ labels in the nuclear responses must be reversed -- these are related to each other in the following way 
\begin{equation}
\begin{split}
    F_{i,j}^{n,p}\propto  c_i^n c_j^{p} F^{n,p}_{X,Y}& \propto c_i^n c_j^{p} \sum_{J=0}^{2J_i} \langle J_i || X_J^{(n)} || J_i \rangle \langle J_i || Y^{(p)}_J || J_i \rangle \\
    F_{j,i}^{n,p}\propto  c_j^{n} c_i^p F^{n,p}_{Y,X}& \propto c_j^n c_i^{p} \sum_{J=0}^{2J_i} \langle J_i || Y_J^{(n)} || J_i \rangle \langle J_i || X^{(p)}_J || J_i \rangle \\
                                                     &\propto c_j^n c_i^{p}F^{p,n}_{X,Y} \\
    F_{i,j}^{p,n}\propto  c_i^p c_j^{n} F^{p,n}_{X,Y} &\propto c_i^p c_j^{n} \sum_{J=0}^{2J_i} \langle J_i || X_J^{(p)} || J_i \rangle \langle J_i || Y^{(n)}_J || J_i \rangle \\
    F_{j,i}^{p,n}\propto  c_j^{p} c_i^n F^{p,n}_{Y,X}&\propto c_j^p c_i^{n} \sum_{J=0}^{2J_i} \langle J_i || Y_J^{(p)} || J_i \rangle \langle J_i || X^{(n)}_J || J_i \rangle \\
                                                    &\propto c_j^{p} c_i^nF^{n,p}_{X,Y}\\
\end{split} 
\end{equation}

Here, the p-n term of $j,i$ is associated with the n-p nuclear form factor. This is important to note as typically the form factors are computed from the OBDME's using a Mathematica package \cite{Anand:2013yka}, which evaluates the form factors $F_{X,Y}^{(N,N')}$ in Table~\ref{tab:form_fact} as they are, i.e., only $F_{\Sigma',\Delta}^{(N,N')}$ and not $F_{\Delta, \Sigma'}^{(N,N')}$. This transformation must be kept in mind when using interaction models where $F_{1,3}$, $F_{4,5}$, $F_{9,8}$, or $F_{11,12}$ are present.

\subsection{Velocity distribution}\label{sec:vel_dist}

The velocity distribution typically assumed for galactic DM is the Standard Halo Model (SHM), where the DM follows a Maxwell Boltzmann distribution, given in the lab (i.e., Earth) reference frame as
\begin{equation}\label{SHMvel}
    f_{\rm SHM}(v)=\frac{1}{(\pi v_0^2)^{3/2}}\exp\left[-\frac{1}{v_0^2}\left(\vec{v}+\vec{v}_E\right)^2\right].
\end{equation}
The velocity of the Earth is given by $\vec{v}_E=\vec{v}_{\odot}+\vec{v}_t$, where
\begin{equation}
\begin{split}
    \vec{v}_{\odot} &= v_{\odot}(0,0,1),\\
    \vec{v}_t &= v_t(\sin 2\pi t,\sin\gamma\cos 2\pi t, \cos \gamma\cos 2\pi t).\\
\end{split}    
\end{equation}
In this frame of reference, the DM velocity is expressed as $\vec{v} =v(\sin\theta \cos\phi, \sin\theta \sin\phi, \cos\theta)$. The parameters employed for this distribution include $\rho_\odot =0.3$ GeV cm$^{-3}$, $v_0 = 220$ km s$^{-1}$, and $v_{\rm{esc}} = 544$ km s$^{-1}$.\\

To allow for easier computation by separating the particle theory and astrophysics contributions, the observation is made that all the terms in the form factor sum are either independent of velocity, or proportional to $v^2$. This allows for the separation of form factors so that
\begin{equation}
    F^{(N,N')}_{i,j}(v,q) = F^{(N,N'),1}_{ij}(q)+v^2 F^{(N,N'),2}_{i,j}(q).
\end{equation}
This allows us to express the velocity integral as a linear combination of two terms:
\begin{equation}
\begin{split}
   g(v_{\rm min})&=\int \frac{f_{\rm lab}(\vec{v})}{v}{\rm d}^3v= \iint_{\mathcal{D}}v~f_{\rm lab}(\vec{v}){\rm d}v~d\Omega,\\   
   h(v_{\rm min})&= \int v f_{\rm lab}(\vec{v}){\rm d}^3v= \iint_{\mathcal{D}}v^3~f_{\rm lab}(\vec{v}){\rm d}v~d\Omega.\\
\end{split}
\label{vel_ints}
\end{equation}
with $\mathcal{D}$ defined as
\begin{equation}
    v>v_{\rm min}(E_R),\quad |\vec{v}+\vec{v}_E|<v_{\rm esc}.
\end{equation}

We can then write the cross section as two terms with the different velocity dependence:
\begin{equation}
\begin{split}
\frac{{\rm d}\sigma_T}{{\rm d}E_R} 
        &= \frac{1}{v^2}\left(\frac{{\rm d}\sigma_T^1}{{\rm d}E_R} + v^2 \frac{{\rm d}\sigma_T^2}{{\rm d}E_R}\right).
\end{split}
\end{equation}
that form prefactors that, aside from $v_{\rm min}$, do not depend on the particle physics DM model in question. 
The benefit of expressing the rate in this way is that it allows for the separate calculation of the astrophysics and particle physics contributions. This makes computation and comparison for different combinations of DM interaction models and velocity distributions significantly easier to perform, as it removes the need to re-evaluate these integrals for every different DM model. 

\section{Accounting for detector effects}
\label{sec:det_effects}
There are a large number of technical and detailed effects that can change a detectors response, that will vary depending on the exact construction of the detector. We will focus here on four main features that most detectors share; multi-isotope targets, energy transformation, resolution, and efficiency.

\subsection{Multi-isotope targets}
Very few detectors are made of completely isotopically pure materials. As each isotope may be impacted differently by the nuclear shell model calculations we use here, we need to make sure they are individually accounted for. This is done by calculating the scattering rate of DM off each isotope and then summing them together based on their predicted abundance $\xi$. For the Si and Ge targets of SuperCDMS, we assume the standard natural abundance of the available isotopes. This is given in Table \ref{Isotope Table}. Thus the total interaction rate for a target is given by:
\begin{equation}
   \frac{{\rm d}R}{{\rm d}E_R} = \sum_{i} \xi_i \frac{{\rm d}R_i}{{\rm d}E_R}. 
\end{equation}

\subsection{Energy transformation}
Direct detection DM experiments do not usually observe the recoil energy directly -- instead it is measured via some other secondary process, typically ionisation, scintillation, or phonon production, and the `observed' energy is reconstructed from this parameter. The mapping of the recoil to observable energy will depend on the nature of the recoil (being electron or nuclear) as there is typically a quenching that occurs for NR (nuclear recoil) events, where not all of the recoil energy is transferred into the observable, meaning the effective energy reconstructed is smaller than the original recoil \cite{LEWIN_1996}. Most experiments will (either explicitly or implicitly) quote their results in units of `electron equivalent keV' (keV$_{\rm ee}$) to reflect this distinction from recoil energy.\\
Because these are the energy units the detector will record a rate in, as well as transforming the energy itself, the observation rate will be different to the interaction rate:
\begin{equation}
    \frac{{\rm d}R}{{\rm d}E_{\rm obs}} = \left. \frac{{\rm d}E_R}{{\rm d}E_{\rm obs}} \frac{{\rm d}R}{{\rm d}E_R}\right|_{E_R(E_{\rm obs})}.
\end{equation}
The observed energy may have additional uncertainty arising from the calibration process. We neglect this for now, and assume that ER (electron recoil) events are perfectly reconstructed to express the mapping between recoil and observed energy.

We here consider phonon detectors like those used by SuperCDMS \cite{supercdmscollaboration2023strategylowmassdarkmatter}. Generally in a simple recoil (in a detector with no potential difference) the total phonon energy will be exactly equal to the recoil energy. However, when detectors are operated at a voltage $V$ (as for SuperCDMS), the Neganov-Trofimov-Luke (NTL) effect \cite{neganov1985ussr,10.1063/1.341976} means additional phonon energy is generated by the movement of charge pairs and the total phonon energy will be
\begin{equation}
    E_{\rm ph} = E_R\left( 1 + \frac{Y(E_R)}{\varepsilon_{\rm eh}} V \right).
\end{equation}
Here, $Y(E_R)$ is the ionisation yield model, a measure of how much of the deposited energy is available to produce electron-hole pairs, and $\varepsilon_{\rm eh}$ is the energy required to create a single electron hole pair. When this is transformed into keV$_{\rm ee}$ by calibrating to electron recoil events (which have a yield of $\sim$1), the observed energy for a nuclear recoil from DM is:
\begin{equation}
    E_{\rm obs} = \frac{\varepsilon_{\rm eh} + Y(E_R)V}{\varepsilon_{\rm eh} + V}E_R.
\end{equation}

\begin{figure}[!h]
    \centering
    \includegraphics[width=0.9\linewidth]{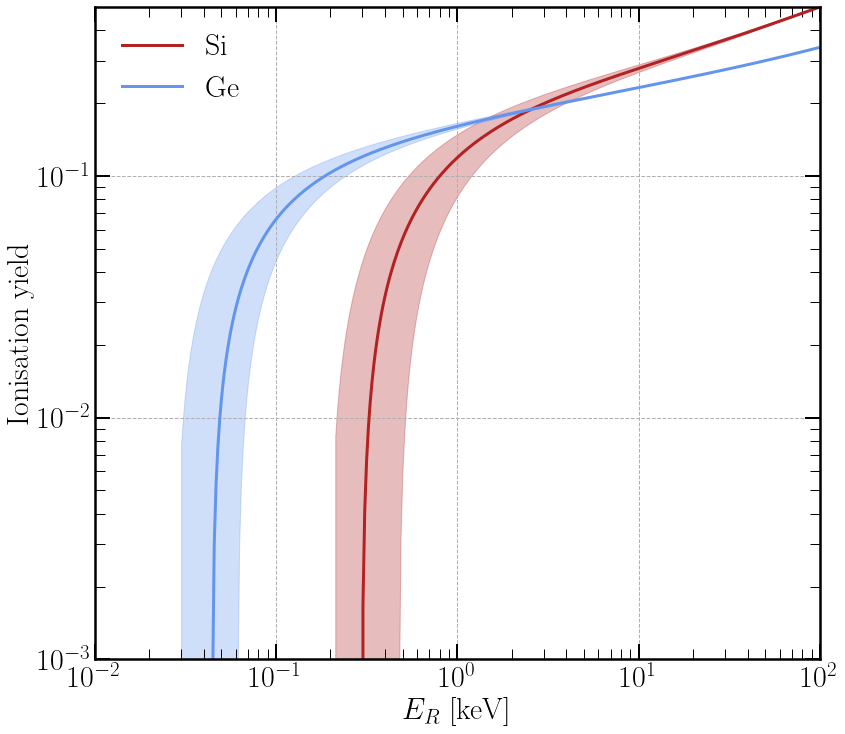}
    \captionsetup{justification=Justified}
    \caption{Ionisation yield for Si and Ge using fit parameters from Ref.~\cite{Sarkis_2020}.}
    \label{fig:ion_yield}
\end{figure}

For this study, we will assume our detectorts are held at 100 V and use the corrected Lindhard model presented in Ref.~\cite{Sarkis_2020}, which is fit to 4 Si data sets and 6 for Ge to construct the curves and shaded uncertainties in Fig.~\ref{fig:ion_yield}.

\subsection{Energy resolution}
DM detectors will have some finite energy resolution, the consideration of which will effectively `smear' any signal observed. An energy deposition that produces a signal of $E_{\rm obs}$ will be transformed by a Gaussian smearing, producing an observed rate of:
\begin{equation}
    \frac{{\rm d}R}{{\rm d}E}=\frac{1}{\sqrt{2\pi}}\int_{0}^{\infty}\frac{1}{\Delta E_{\rm obs}}\frac{{\rm d}R}{{\rm d}E_{\rm obs}}\exp\left[\frac{-(E-E_{\rm obs})^2}{2(\Delta E_{\rm obs})^2}\right]{\rm d}E_{\rm obs},
\end{equation}
where in general the resolution term takes the form:
\begin{equation}
    \Delta E_{\rm obs} = \sqrt{BE_{\rm obs}+(AE_{\rm obs})^2 + \sigma_E^2}.
\label{eq:res}
\end{equation}
The constants in Eq.~\ref{eq:res} are typically found from the calibration of a detector -- they are fit parameters that will have some associated uncertainty.

\subsection{Detector efficiency}
Finally, detectors will have some finite efficiency, typically a function of energy, accounted for by multiplying the rate by the efficiency factor $\epsilon(E)$
\begin{equation}
    \frac{{\rm d}R}{{\rm d}E}=\frac{\epsilon(E)}{\sqrt{2\pi}}\int_{0}^{\infty}\frac{1}{\Delta E_{\rm obs}}\frac{{\rm d}R}{{\rm d}E_{\rm obs}}\exp\left[\frac{-(E-E_{\rm obs})^2}{2(\Delta E_{\rm obs})^2}\right]{\rm d}E_{\rm obs}.
\label{eq:full_rate}
\end{equation}
This function tends to be reported by an experiment for a given dataset and analysis. 

For simplicity, and to isolate the impact of the nuclear structure uncertainty, we will assume no additional uncertainty from these detector effects. For our toy SuperCDMS-like detectors considered here, we make the assumptions given in Table \ref{tab:exp_params} for all following calculations.

\begin{table}[htpb]
\captionsetup{justification=Justified}
\caption{Experimental parameters assumed for Ge and Si detectors.\label{tab:exp_params}}
\centering
\begin{tabular}{|c|c|c|}
\hline
Parameter & Ge & Si \\ \hline
$A_E$ (unitless) & $5\times 10^{-3}$ & $5\times 10^{-3}$ \\
$B_E$ (eV) & $0.7$ & $0.7$ \\
$\sigma_E$ (eV) & $10$ & $13$ \\
$Y(E_R)$ & \cite{Sarkis_2020}, 100V & \cite{si_yield}, 100V \\
$\epsilon(E)$ & 85\% (flat) & 85\% (flat) \\
Energy range & $0.1-10$ keV & $0.1-10$ keV \\
$R_b$ (cpd/kg/keV) & $3\times 10^{-3}$ & 0.1 \\
Exposure (kg$\times$ yrs) & 44.5 & 9.6 \\ \hline
\end{tabular}
\end{table}

\section{Uncertainties in nuclear structure}
\label{sec:nuclear_struc}

The OBDMEs used in these calculations are obtained from large-scale nuclear shell model calculations using NuShellX \cite{Brown:2014bhl} for the silicon  $^{28,29,30}$Si and germanium $^{70,72,73,74,76}$Ge isotopes. Table~\ref{Isotope Table} summarises these isotopes, alongside their natural abundance and ground state spin-parity values $J_i^\pi$. 

\begin{table*}[htpb] 
\captionsetup{justification=Justified}
\caption{Silicon and germanium isotopes considered in this work, alongside the natural abundance values and ground state spin-parity $J_i^\pi$. Shell model interactions employed are displayed, alongside the valence space levels and valence space truncation (unrestricted).}
\centering
\begin{tabular}{|p{3.2cm}|c|c|c|c|c|c|}
 \hline
  Nucleus & Z  & A & Abundance & $J_i^\pi$ & Valence Space & Shell Model Interactions\\
 \hline
  \multirow{3}{5mm}{\text{Silicon (Si)}} & \multirow{3}{5mm}{14}  & 28 & 92.2\% & $0^+$ & $sd$ & \multirow{3}{25mm}{\text{- USD \cite{Brown:1988vm, Wildenthal:1984mf}} \text{- USDB \cite{Brown:2006gx}}} \\
  &  & 29 & 4.7\% & $1/2^+$ & \text{$1d_{5/2}$, $2s_{1/2}$, $1d_{3/2}$} &\\
  &  & 30 & 3.1\% & $0^+$ & Unrestricted &\\
  \hline
  \multirow{5}{5mm}{\text{Germanium (Ge)}} & \multirow{5}{5mm}{32}  & 70 & 20.57\% & $0^+$ &  & \multirow{5}{25mm}{\text{- GCN2850 \cite{Menendez:2008jp}} \text{- JUN45 \cite{Honma:2009zz}} \text{- jj44b \cite{Mukhopadhyay:2017tca}}}\\
   &  & 72 & 27.45\% & $0^+$ & $f_5pg_9$ &\\
   &  & 73 & 7.75\% & $9/2^+$ & $2p_{3/2}$,  $1f_{5/2}$, $2p_{1/2}$, $1g_{9/2}$ &\\
   &  & 74 & 36.50\% & $0^+$ & Unrestricted (JUN45, jj44b) &\\
   &  & 76 & 7.73\% & $0^+$ & &\\
  \hline 
\end{tabular}
\label{Isotope Table}
\end{table*}
 
The spherical shell model employed for a nuclear basis assumes an inert core, made up of completely filled levels and shells, which has spin-parity $J_i^\pi = 0^+$. The shell above this core is termed the valence (model) space, and constitutes the valence nucleons whose configuration across the valence single-particle levels dictates the nuclear wave function. All levels and shells above this valence space are taken to be empty. Nuclear shell model programs, like NuShellX \cite{Brown:2014bhl}, compute these configurations, given a choice of shell model interaction and valence space truncation (limits imposed on the occupation numbers of certain valence space levels). The OBDMEs (and hence form factors) calculated using NuShellX will differ depending on the shell model interaction employed, introducing an uncertainty that may propagate to WIMP interaction rates and ultimately experimental sensitivities. 

A summary of the valence space levels and shell model interactions employed for each nuclear target is provided in Table~\ref{Isotope Table}. In this work, calculations performed using NuShellX (all except GCN2850) use an unrestricted valence space, with no valence space truncation imposed. Experimental nuclear observables were compared to the theoretical NuShellX counterparts in Refs.~\cite{AbdelKhaleq:2023ipt,AbdelKhaleq:2024hir}, as a measure of the accuracy of the computational shell model predictions, in particular of the ground state wave function. Further details of this comparison can be found in these works -- we omit them here for brevity.

The silicon calculations employ the USD \cite{Brown:1988vm, Wildenthal:1984mf} and USDB \cite{Brown:2006gx} shell model interactions in an $sd$ model space. The Hamiltonian of the former was obtained from a least-squares fit to 380 energy states from 66 nuclei with $16\leq A\leq40$, whilst the latter Hamiltonian is based on fittings to the energies of 608 states of 77 nuclei with $21\leq A\leq40$ \cite{Brown:2006gx}. The GCN2850 \cite{Menendez:2008jp} form factors employed for the germanium analysis were obtained from the Fitzpatrick \textit{et al}. \cite{Fitzpatrick:2012ix,Anand:2013yka} work, with a valence space truncation limiting the $1g_{9/2}$ level occupation number to no more than two nucleons above the minimum occupation for all isotopes. Additionally,  OBDME calculations were performed \cite{AbdelKhaleq:2023ipt} using the JUN45 \cite{Honma:2009zz} and jj44b \cite{Mukhopadhyay:2017tca} shell model interactions in an $f_5pg_9$ valence space. The JUN45 interaction was fitted to 400 experimental binding energy and excitation energy data for 69 nuclei with $63 \leq A \leq 96$; jj44b was fitted using 77 binding energy and 470 excitation energy data for nuclei with $Z = 28-30$ ($N = 28-50$) and $N = 48-50$ ($Z = 28-50$). Additional details on the aforementioned shell model interactions can be found in the corresponding associated papers.

An integrated form factor (IFF) value was employed in Ref.~\cite{AbdelKhaleq:2023ipt} as a gauge of the strength of each of the nuclear channels $M$, $\Sigma''$, $\Sigma'$, $\Delta$, $\Phi''$, and $\tilde{\Phi}'$, allowing for a comparison of their behaviour under different assumptions. These IFFs are given by

\begin{equation}\label{IFF expressions}
  \int_{0}^{q_{\rm max}} \frac{q \,dq}{2} F^{(N, N)}_{X(,Y)} (q^2),
\end{equation}

\noindent in units of $\text{MeV}^2$. We note that the choice of $q_{\rm max}$ may impact the nuclear uncertainty present between the IFFs of two different shell model calculations for a particular nucleus. Here, we perform IFF calculations using a $q$ range that better reflects the experimental energy analysis region, which corresponds to an upper limit of $q_{\rm max} =50$ MeV for both silicon and germanium. Although all results are presented for this $q_{\rm max}$ value, in Appendix~\ref{appendix: IFFs 100 MeV} we present a brief comparison between this range and the $q_{\rm max}=100$ MeV values employed in previous works \cite{Fitzpatrick:2012ix,Anand:2013yka,AbdelKhaleq:2023ipt}.

Given that the cross section $\sigma_T$ in Eq.~\ref{full_cross} is written in terms of the $F_{i,j}^{(N,N')}$ nuclear responses (Table~\ref{tab:form_fact}) instead of the $F_{X,Y}^{(N,N')}$ terms, it is more convenient to redefine the IFF integral in order to better gauge the nuclear uncertainty propagation from the nuclear form factors to the cross section. The IFF value is now evaluated through

\begin{equation}
    {\rm IFF}_{i,j}^{(N,N')} (v,q_{\rm max},B) = \int_0^{q_{\rm max}} \frac{q dq}{2} F_{i,j}^{(N,N')} (q,v,B),
\end{equation}

\noindent where $B$ designates the shell model interaction employed in the calculation. Here, this analysis is performed for each set of $i,i$ in $F^{(N,N')}_{ij}(v,q,B)$ separately, where only the $q$-dependent terms in each response are integrated over. Based on Table~\ref{tab:form_fact}, the terms we need to account for in this integral are

\begin{equation}\label{eq:IFFij}
\begin{split}
    & F_{M},  \hspace{5mm} F_{\Sigma''}, \hspace{5mm} F_{\Sigma'},\\
    &  \frac{q^2}{m_N^2} F_{\Sigma'}, \hspace{5mm} \frac{q^2}{m_N^2} F_{M}, \hspace{5mm} \frac{q^2}{m_N^2} F_{\Delta}, \hspace{5mm} \frac{q^2}{m_N^2} F_{\Sigma''},  \\
    & \frac{q^2}{m_N^2} F_{\Phi''}, \hspace{5mm} \frac{q^2}{m_N^2} F_{\tilde{\Phi}'}, \hspace{5mm} \frac{q^2}{m_N^2} F_{\Sigma',\Delta}, \hspace{5mm} \frac{q^2}{m_N^2} F_{M, \Phi''}, \\
    & \frac{q^4}{m_N^4} F_{\Phi''}, \hspace{5mm} \frac{q^4}{m_N^4} F_{\Delta}, \hspace{5mm} \frac{q^4}{m_N^4} F_{\Sigma''}.
\end{split}
\end{equation}

From here on we denote $k\equiv q/m_N$. The factors of $1/m_N$ have been kept in the IFF expressions and analysis, as they help retain the relative magnitude between two IFF nuclear channels, in line with the cross section expression. This allows for better understanding of which dominant IFF channels and their corresponding nuclear uncertainties are expected to propagate to the DM experimental limits. The IFF plots are separated into terms which depend on a sum over even $J$ values ($ F_{M}$, $k^2 F_{M}$, $k^2 F_{\Phi''}$, $k^2 F_{\tilde{\Phi}'}$, $k^2 F_{M, \Phi''}$, $k^4 F_{\Phi''}$) and odd $J$ values ($F_{\Sigma''}$,  $F_{\Sigma'}$, $k^2 F_{\Sigma''}$, $k^2 F_{\Sigma'}$, $k^2 F_{\Delta}$, $k^2 F_{\Sigma',\Delta}$, $k^4 F_{\Sigma''}$, $k^4 F_{\Delta}$). Values on top of IFF bars indicate factor differences between the shell model calculations, with the arrows indicating direction of multiplication.

\begin{figure*}[htpb]
    \centering
    \includegraphics[width=0.3573\textwidth]{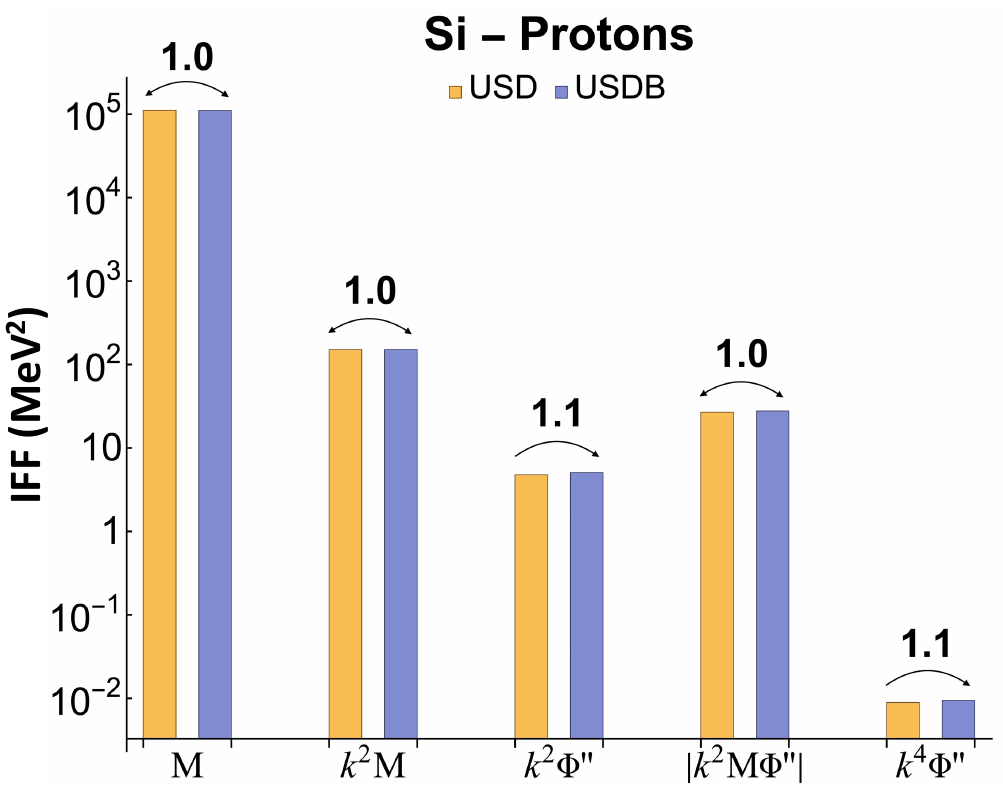}~
    \includegraphics[width=0.3573\textwidth]{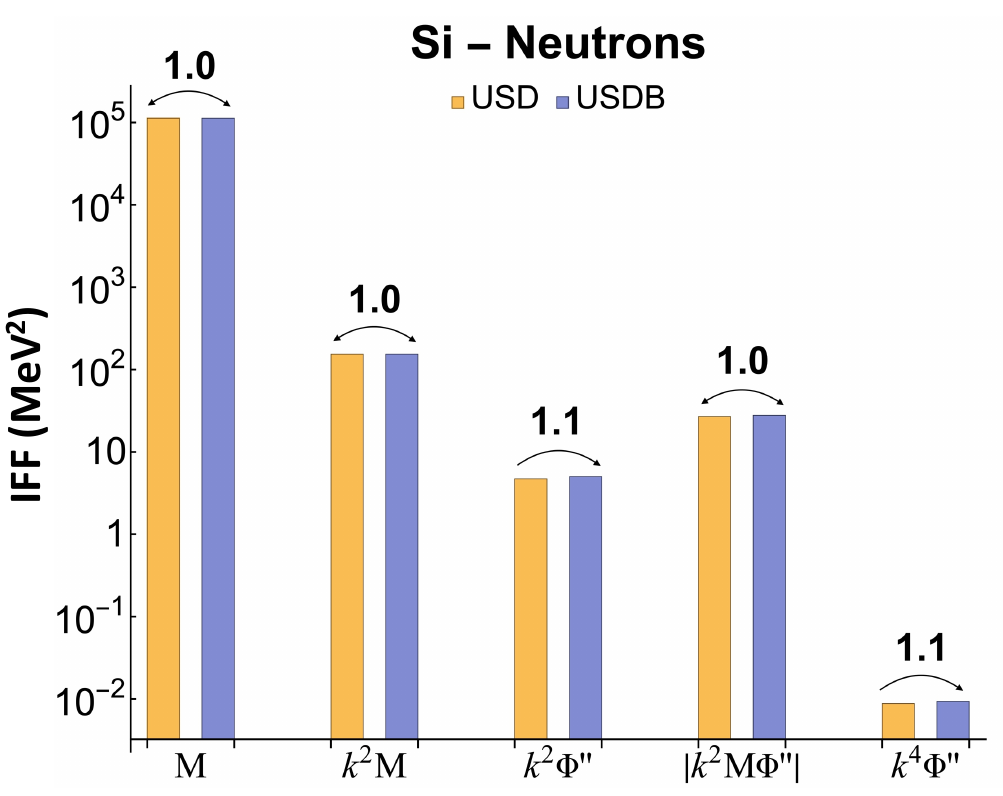}\\
    \includegraphics[width=0.3573\textwidth]{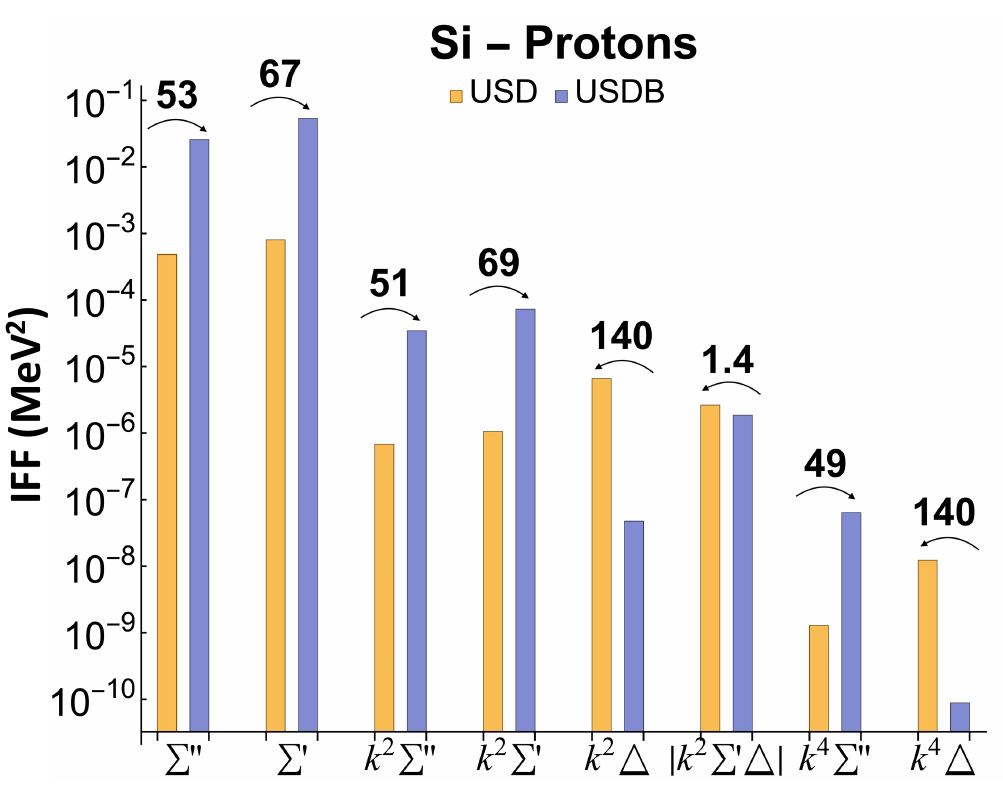}~
    \includegraphics[width=0.3573\textwidth]{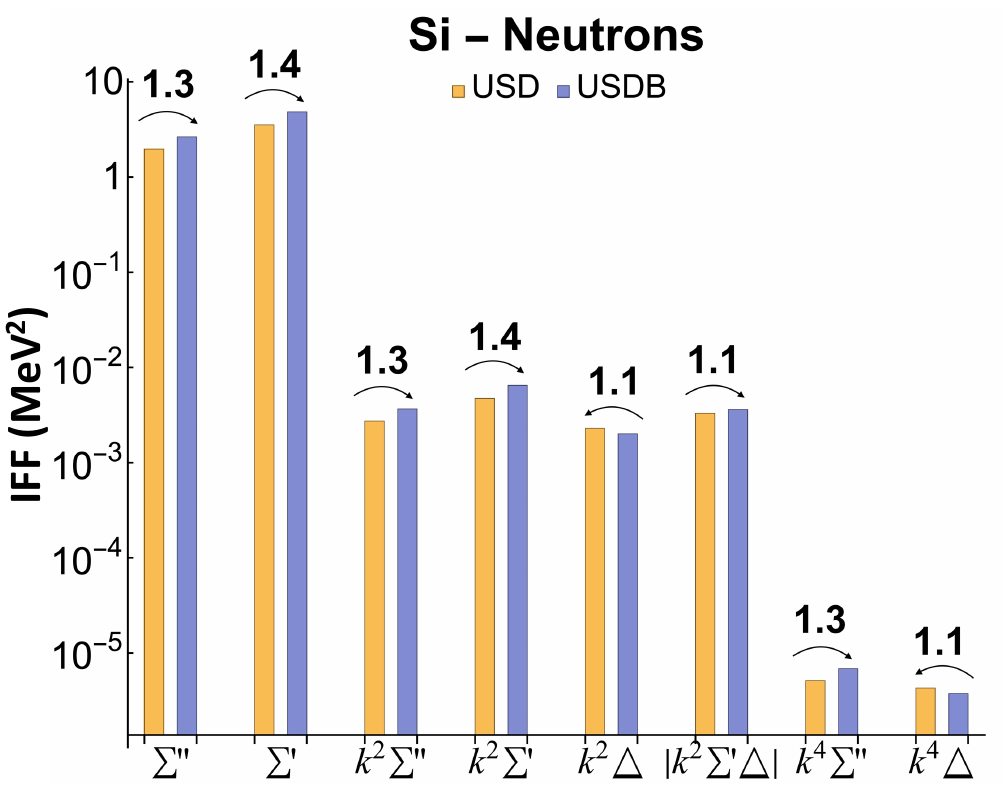}
    \captionsetup{justification=Justified}
    \caption{Silicon proton (left) and neutron (right) IFF values for an integral with $q_{\rm max}= 50$ MeV. Top panel shows values related to the nuclear responses $M$, $\Phi''$, whilst bottom panel shows those related to $\Sigma''$, $\Sigma'$, $\Delta$.}
    \label{fig: silicon 50 MeV IFFs}
\end{figure*}

\begin{figure*}[htpb]
    \centering
    \includegraphics[width=0.3573\textwidth]{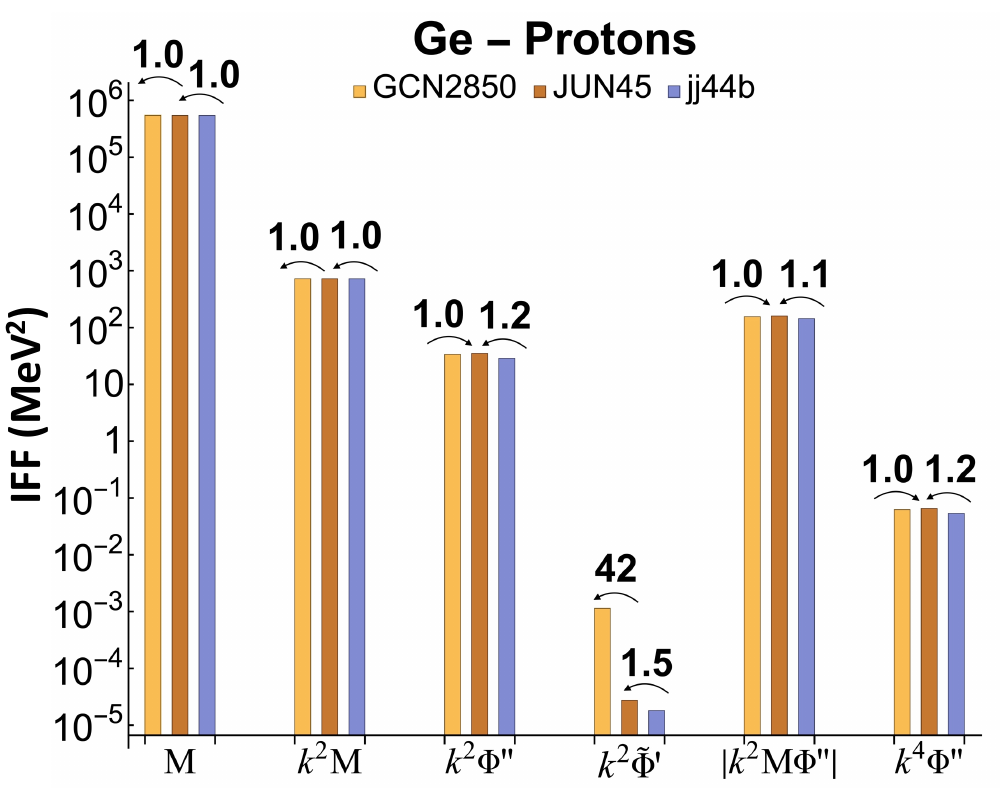}~
    \includegraphics[width=0.3573\textwidth]{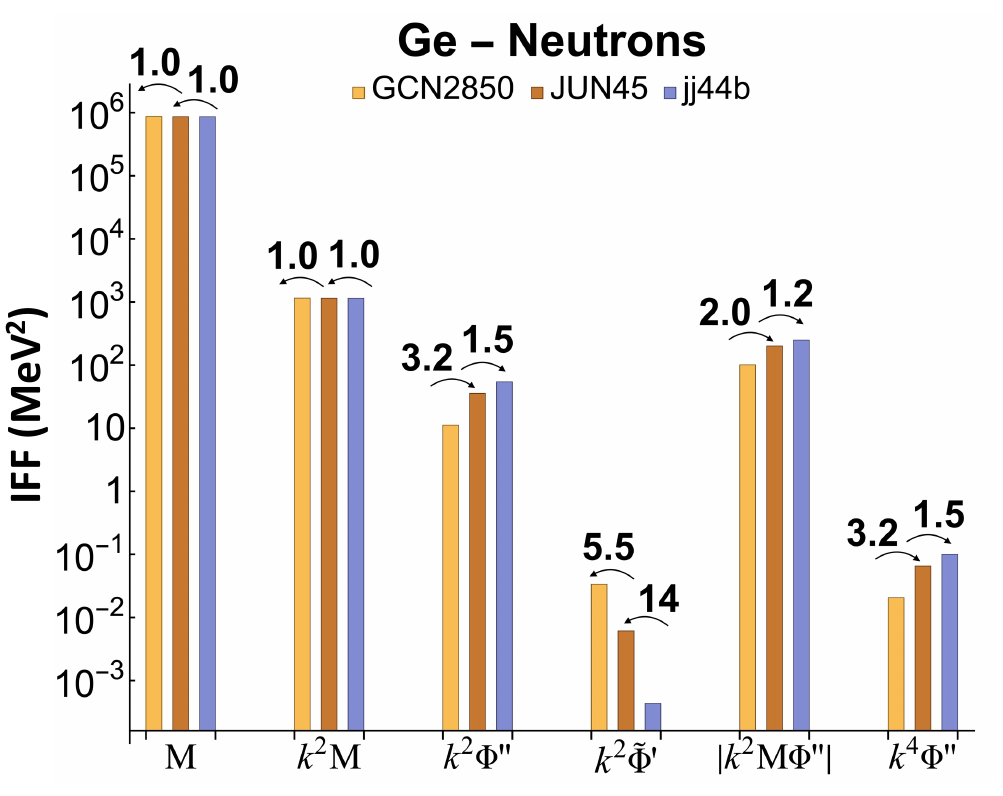}\\
    \includegraphics[width=0.3573\textwidth]{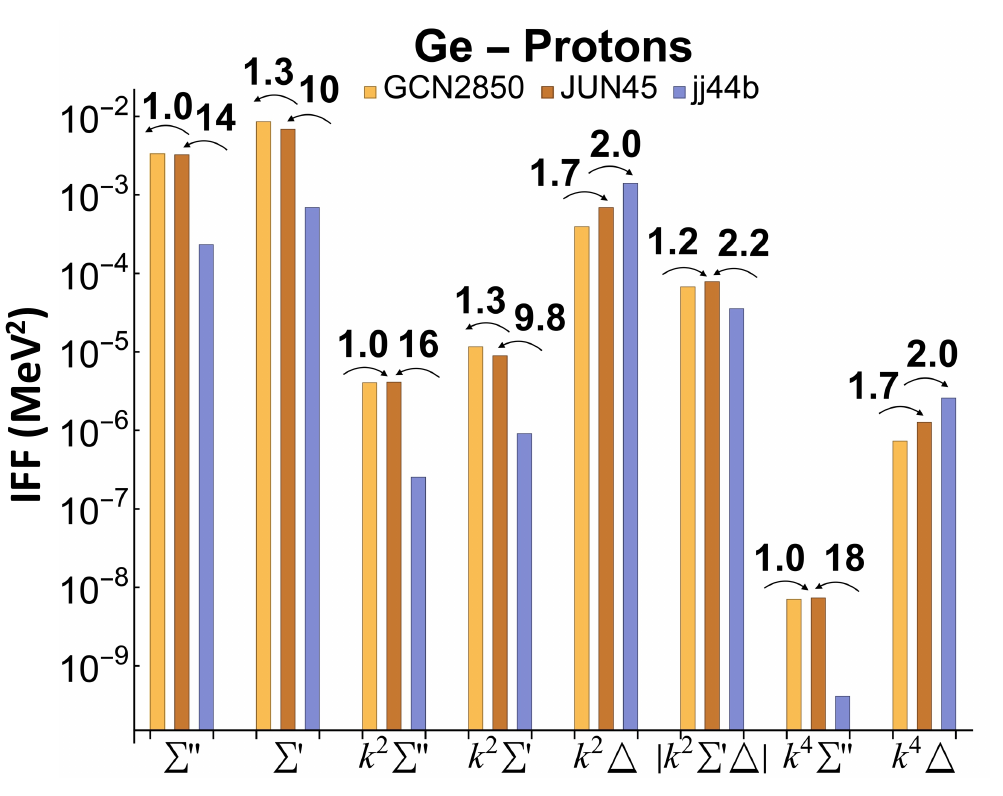}~
    \includegraphics[width=0.3573\textwidth]{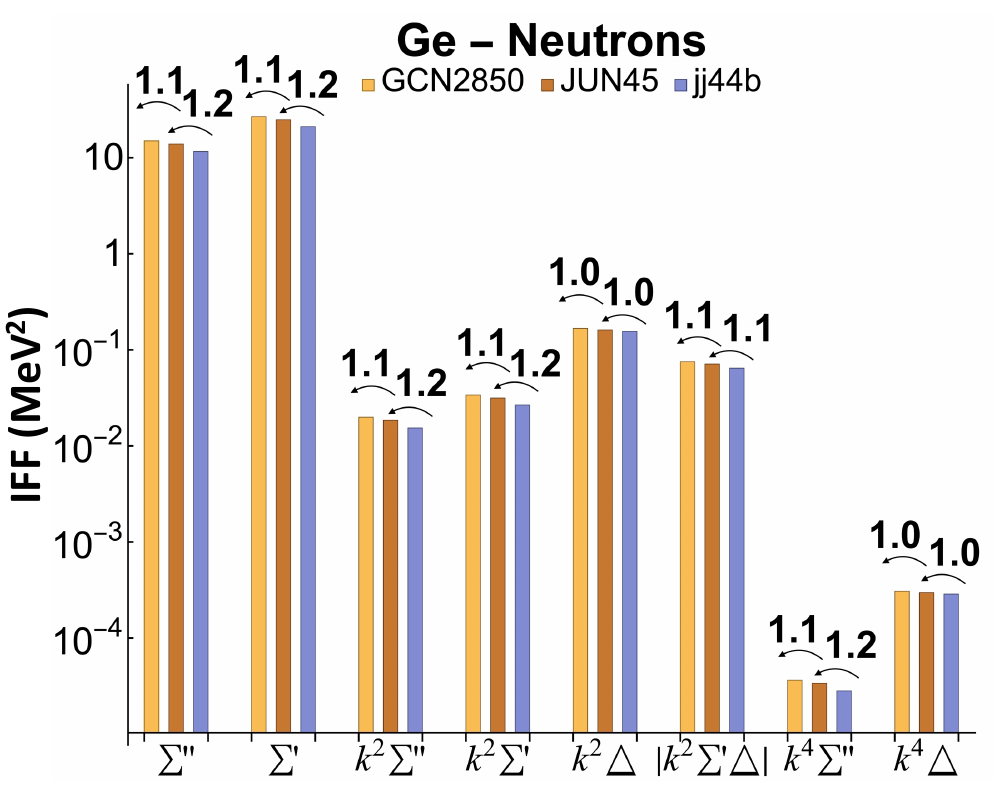}
    \captionsetup{justification=Justified}
    \caption{Germanium proton (left) and neutron (right) IFF values for an integral with $q_{\rm max}= 50$ MeV. Top panel shows values related to the nuclear responses $M$, $\Phi''$, $\tilde{\Phi}'$, whilst bottom panel shows those related to $\Sigma''$, $\Sigma'$, $\Delta$.}
    \label{fig: germanium 50 MeV IFFs}
\end{figure*}

These $q_{\rm max}= 50$ MeV IFFs are are presented for silicon and germanium in Figs.~\ref{fig: silicon 50 MeV IFFs} and \ref{fig: germanium 50 MeV IFFs}, respectively. As was done previously in Ref.~\cite{AbdelKhaleq:2023ipt}, the IFFs are decomposed into the proton-proton and neutron-neutron components, in order to gauge the relative strength and contribution of each nucleon type. Table~\ref{tab:ratios_table} provides a summary of the IFF nuclear uncertainties, alongside the corresponding $\mathcal{O}_{i}^{\rm NR}$ responses.

As expected, the SI $M$ IFF values stay consistent across shell model calculations for both silicon and germanium. The nuclear uncertainties between $q$-independent and $q$-dependent channels remain mostly consistent, indicating that this additional dependence does not impact the nuclear uncertainties for each channel $X$. Nuclear channels with larger factors of $k\equiv q/m_N$ are further suppressed in magnitude, potentially impacting their nuclear uncertainty propagation to the DM exclusion limits.

\section{Experimental limits}
\label{sec:results}

To understand the full impact of considering different nuclear structure models for each NREFT operator, we consider three different types of particle interactions:
\begin{enumerate}
    \item Stronger proton couplings to DM ($\hat{c}_i^n=0.01\hat{c}_i^p$)\\
    \item Stronger neutron couplings to DM ($\hat{c}_i^p=0.01\hat{c}_i^n$)\\
    \item The $\hat{c}_i^n/\hat{c}_i^p$ ratio derived from assuming the quark level couplings for the related high energy operator in Appendix \ref{sec:quark_couplings}.
\end{enumerate}
For those that do not have high energy counterparts, we assume equal couplings ($\hat{c}_i^p=\hat{c}_i^n$). These ratios are shown in Table \ref{tab:coupling_ratios}. Where a model depends on the DM spin, we take $j_{\chi}=1/2$. We note that in this high energy mapping, each NR operator $\mathcal{O}_{i}^{\rm NR}$ can be mapped to from more than one relativistic operator $\mathcal{O}_j$. All such mappings are presented in Table \ref{tab:coupling_ratios}, however where more than one $\hat{c}_i^n/\hat{c}_i^p$ exists we employ the one where $\hat{c}_i^n/\hat{c}_i^p \neq 1$ in our analysis and discussion. We note that only the coefficient ratios are obtained from this high energy mapping -- the energy scale of the new physics $\Lambda$ is absorbed into the definition of the cross section being constrained. 

\begin{table}[!h]
\captionsetup{justification=Justified}
\caption{Neutron to proton coupling ratio computed from quark level couplings. }
\label{tab:coupling_ratios}
\begin{tabular}{@{}lll@{}}
\toprule
\textbf{NR operator} & \textbf{Relativistic operator} & $\hat{c}_i^n/\hat{c}_i^p$ \\ \midrule
\ONR{1} & $\mathcal{O}_1$, $\mathcal{O}_5$ & 1, 1 \\
\ONR{3} & - & 1 \\
\ONR{4} & $\mathcal{O}_8$, $\mathcal{O}_9$ & 1, 1 \\
\ONR{5} & - & 1 \\
\ONR{6} & $\mathcal{O}_4$ & -0.19 \\
\ONR{7} & $\mathcal{O}_7$ & 1 \\
\ONR{8} & $\mathcal{O}_6$ & 1 \\
\ONR{9} & $\mathcal{O}_6$, $\mathcal{O}_7$ & 1, 1\\
\ONR{10} & $\mathcal{O}_3$, $\mathcal{O}_{10}$ & -0.19, 1 \\
\ONR{11} & $\mathcal{O}_2$, $\mathcal{O}_{10}$ & 1, 1 \\
\ONR{12} & $\mathcal{O}_{10}$ & 1 \\ \bottomrule
\end{tabular}
\end{table}

\begin{figure*}[htpb]
    \centering
    \includegraphics[width=\linewidth]{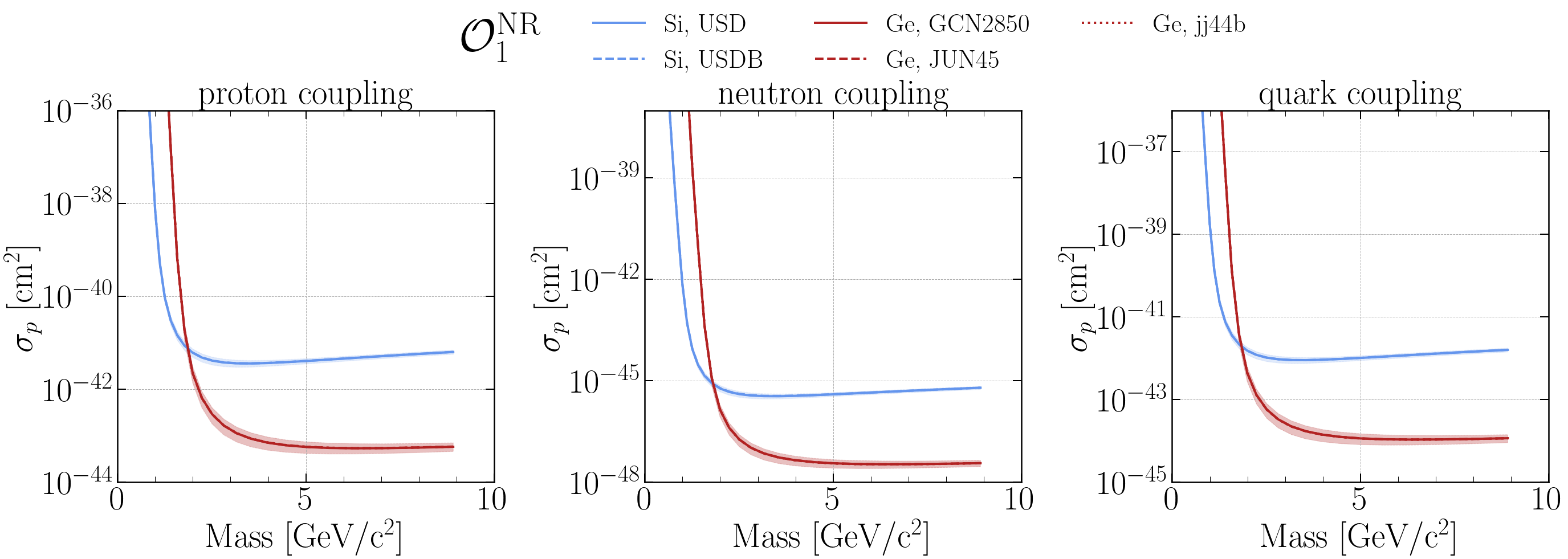}
    \captionsetup{justification=Justified}
    \caption{Limits on the effective DM-proton cross section for \ONR{1}. Proton coupling refers to the scenario where $\hat{c}_i^n=0.01\hat{c}_i^p$ while neutron coupling takes $\hat{c}_i^p=0.01\hat{c}_i^n$. The quark coupling case uses the ratios given in Table \ref{tab:coupling_ratios}.}
    \label{fig:O1_limits}
\end{figure*}

\begin{figure*}[htpb]
    \centering
    \includegraphics[width=\linewidth]{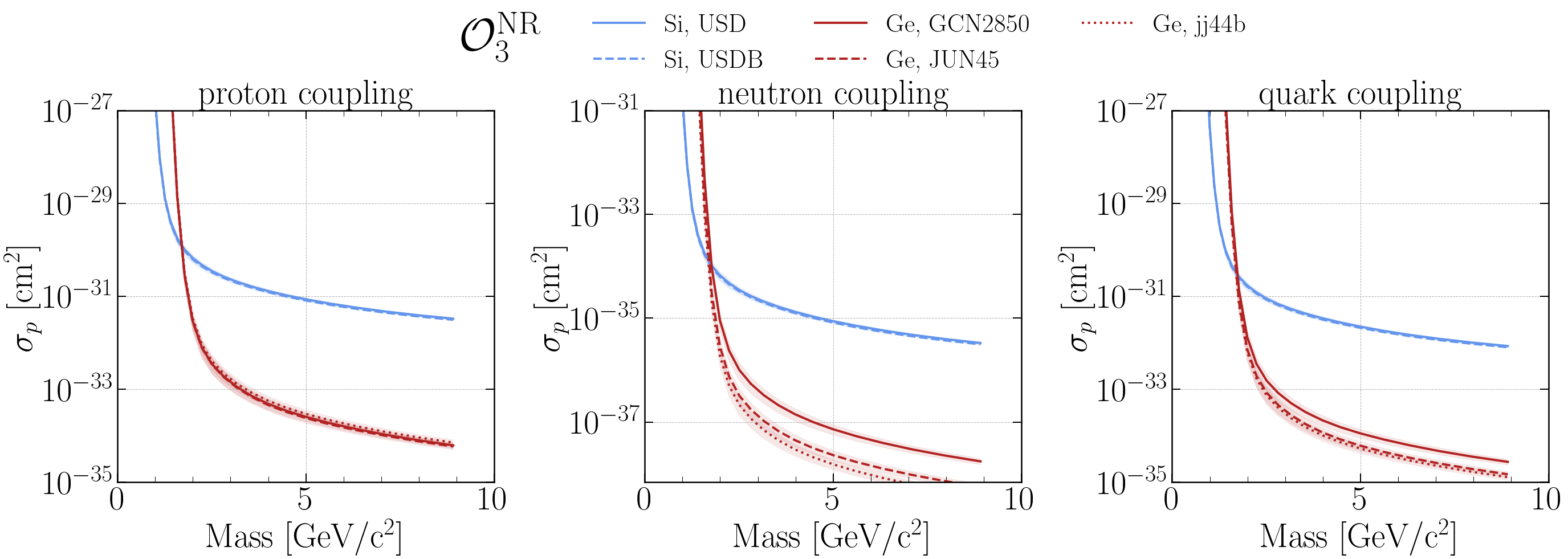}
    \captionsetup{justification=Justified}
    \caption{Limits on the effective DM-proton cross section for \ONR{3}. Proton coupling refers to the scenario where $\hat{c}_i^n=0.01\hat{c}_i^p$ while neutron coupling takes $\hat{c}_i^p=0.01\hat{c}_i^n$. The quark coupling case uses the ratios given in Table \ref{tab:coupling_ratios}.}
    \label{fig:O3_limits}
\end{figure*}

\begin{figure*}[htpb]
    \centering
    \includegraphics[width=\linewidth]{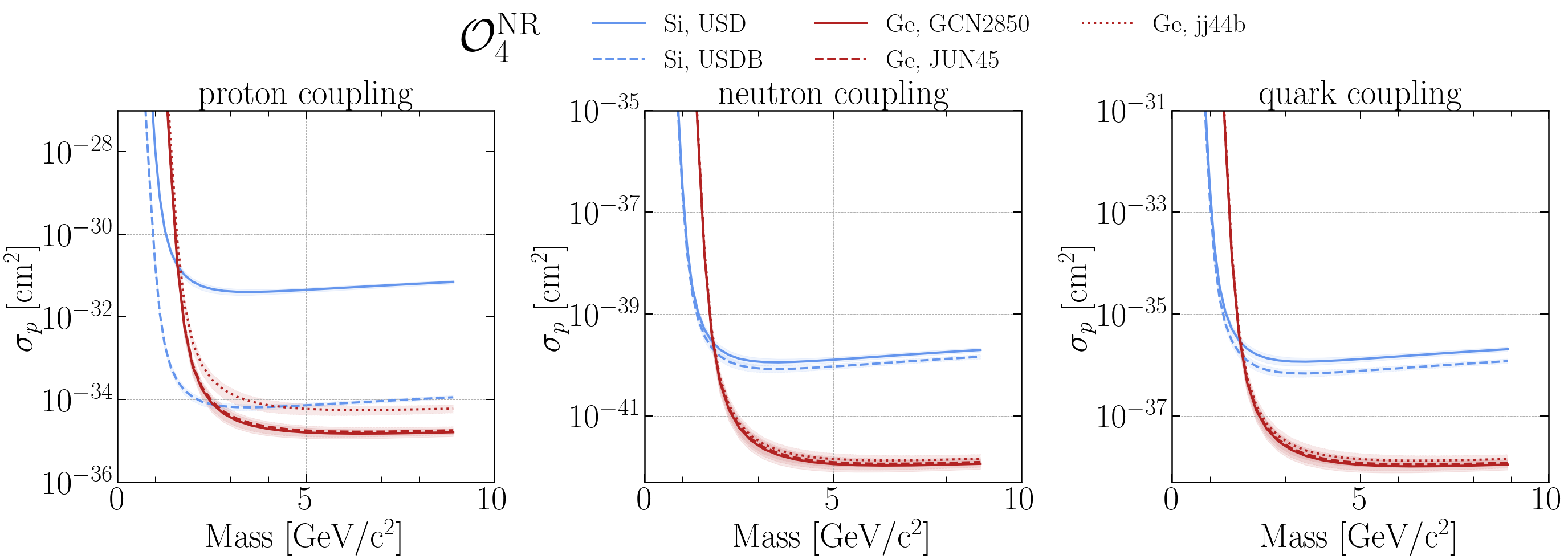}
    \captionsetup{justification=Justified}
    \caption{Limits on the effective DM-proton cross section for \ONR{4}. Proton coupling refers to the scenario where $\hat{c}_i^n=0.01\hat{c}_i^p$ while neutron coupling takes $\hat{c}_i^p=0.01\hat{c}_i^n$. The quark coupling case uses the ratios given in Table \ref{tab:coupling_ratios}.}
    \label{fig:O4_limits}
\end{figure*}

\begin{figure*}[htpb]
    \centering
    \includegraphics[width=\linewidth]{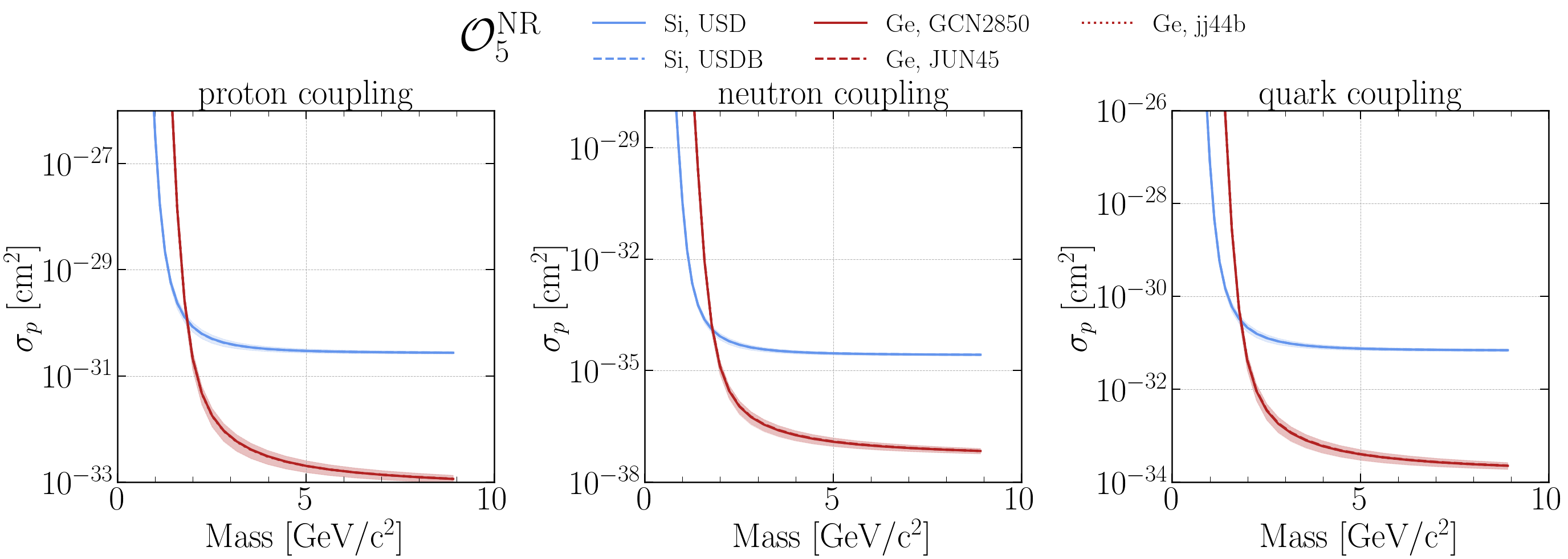}
    \captionsetup{justification=Justified}
    \caption{Limits on the effective DM-proton cross section for \ONR{5}. Proton coupling refers to the scenario where $\hat{c}_i^n=0.01\hat{c}_i^p$ while neutron coupling takes $\hat{c}_i^p=0.01\hat{c}_i^n$. The quark coupling case uses the ratios given in Table \ref{tab:coupling_ratios}.}
    \label{fig:O5_limits}
\end{figure*}

\begin{figure*}[htpb]
    \centering
    \includegraphics[width=\linewidth]{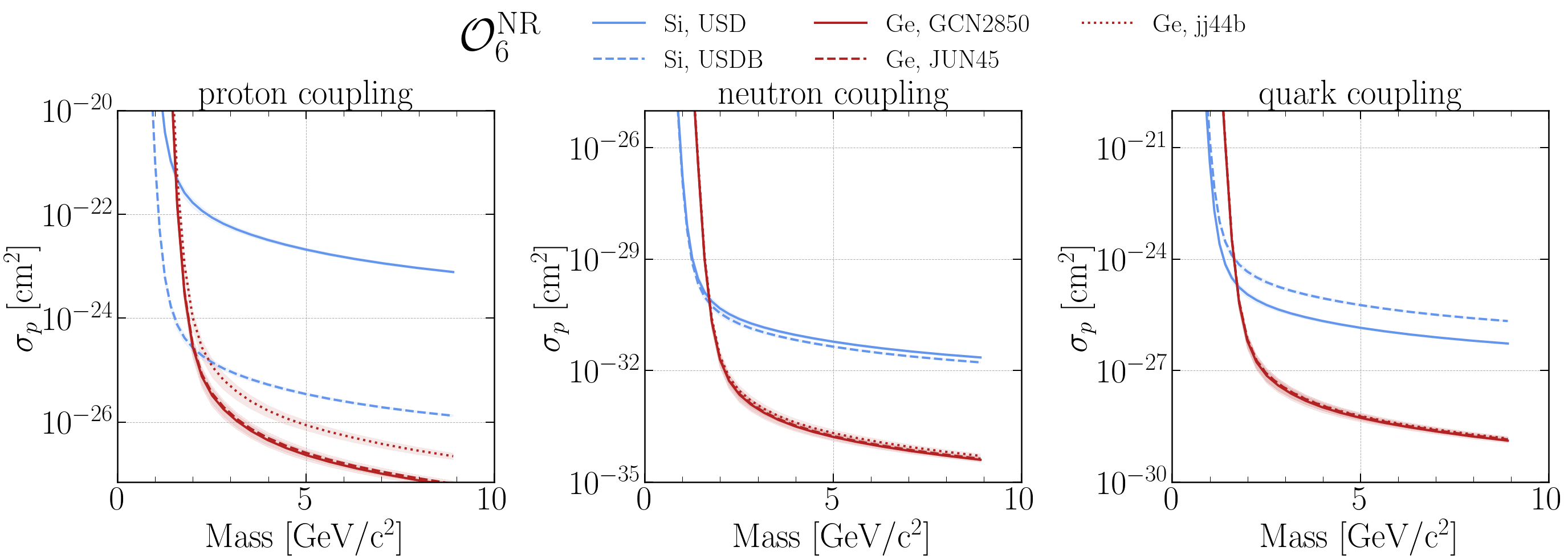}
    \captionsetup{justification=Justified}
    \caption{Limits on the effective DM-proton cross section for \ONR{6}. Proton coupling refers to the scenario where $\hat{c}_i^n=0.01\hat{c}_i^p$ while neutron coupling takes $\hat{c}_i^p=0.01\hat{c}_i^n$. The quark coupling case uses the ratios given in Table \ref{tab:coupling_ratios}.}
    \label{fig:O6_limits}
\end{figure*}

\begin{figure*}[htpb]
    \centering
    \includegraphics[width=\linewidth]{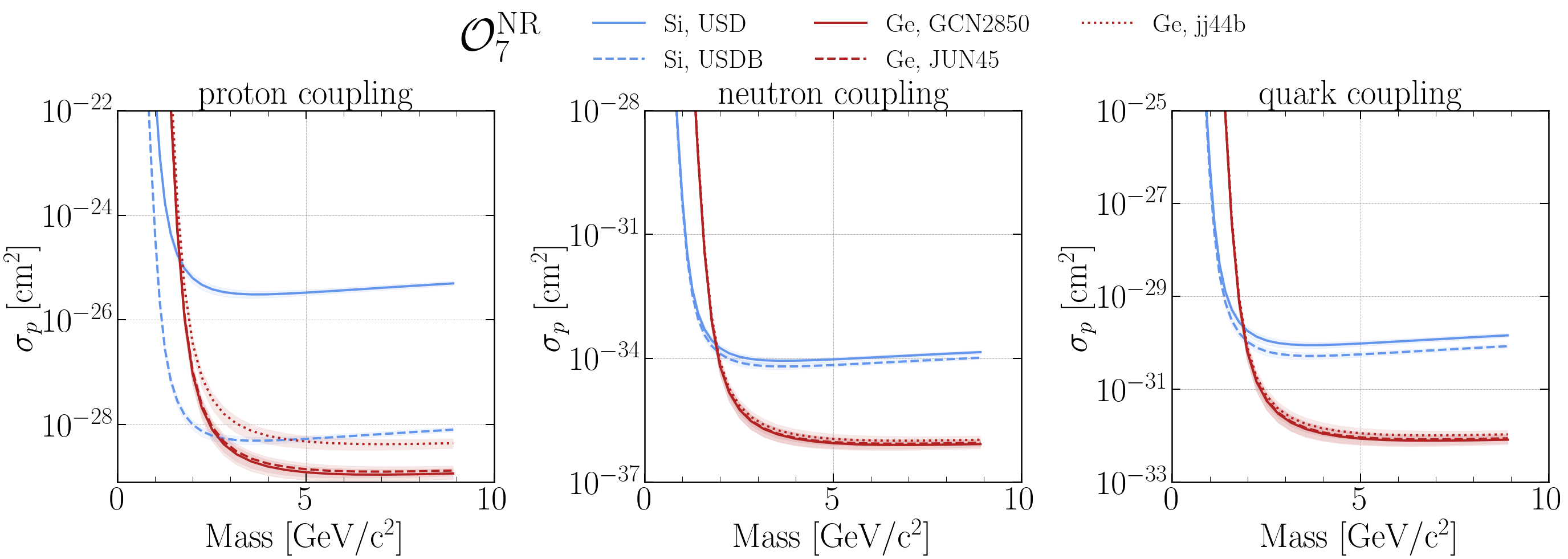}
    \captionsetup{justification=Justified}
    \caption{Limits on the effective DM-proton cross section for \ONR{7}. Proton coupling refers to the scenario where $\hat{c}_i^n=0.01\hat{c}_i^p$ while neutron coupling takes $\hat{c}_i^p=0.01\hat{c}_i^n$. The quark coupling case uses the ratios given in Table \ref{tab:coupling_ratios}.}
    \label{fig:O7_limits}
\end{figure*}

\begin{figure*}[htpb]
    \centering
    \includegraphics[width=\linewidth]{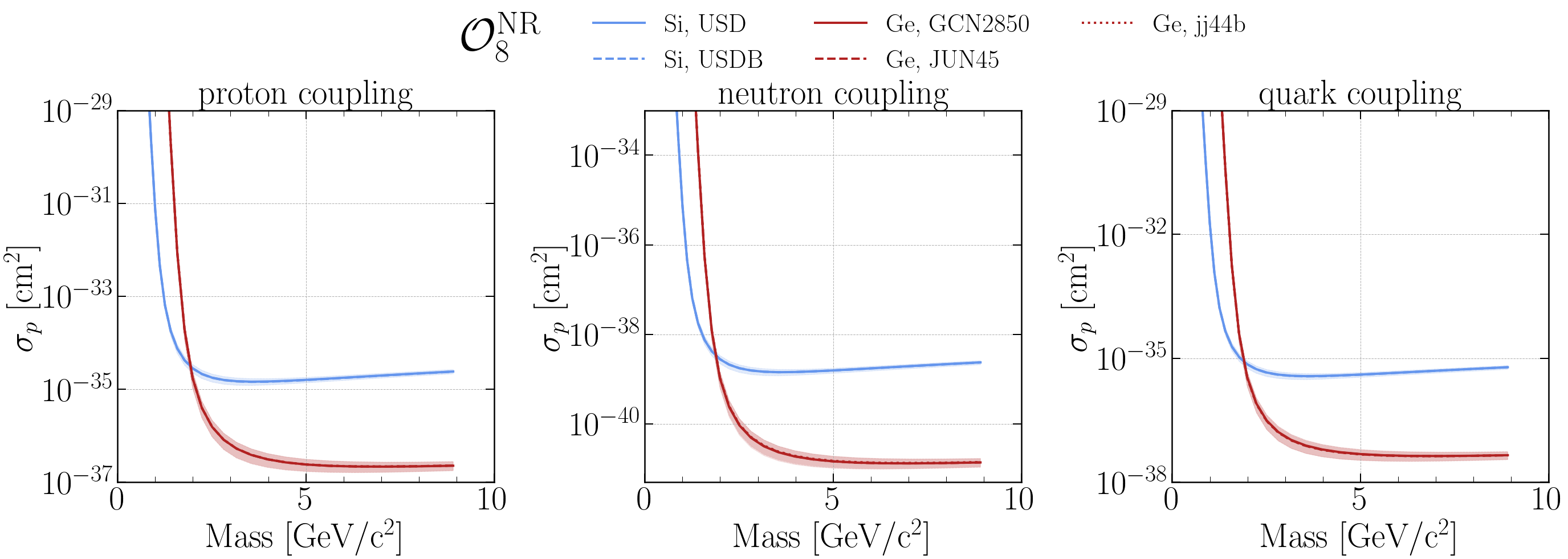}
    \captionsetup{justification=Justified}
    \caption{Limits on the effective DM-proton cross section for \ONR{8}. Proton coupling refers to the scenario where $\hat{c}_i^n=0.01\hat{c}_i^p$ while neutron coupling takes $\hat{c}_i^p=0.01\hat{c}_i^n$. The quark coupling case uses the ratios given in Table \ref{tab:coupling_ratios}.}
    \label{fig:O8_limits}
\end{figure*}

\begin{figure*}[htpb]
    \centering
    \includegraphics[width=\linewidth]{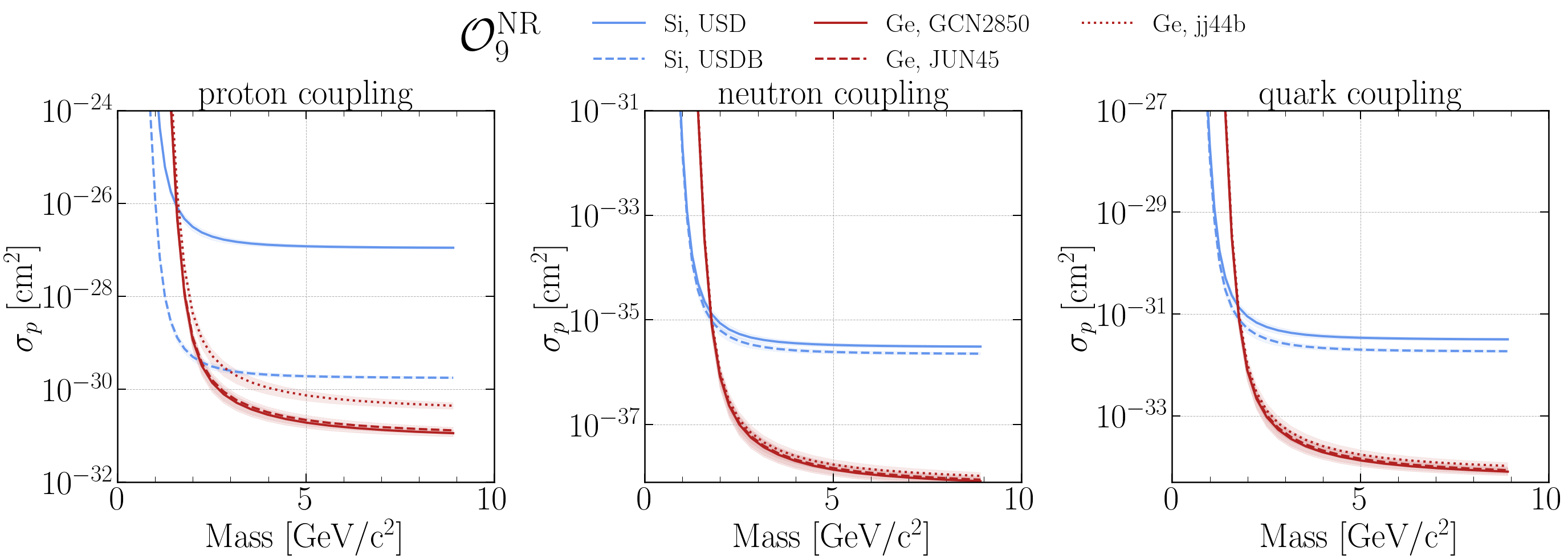}
    \captionsetup{justification=Justified}
    \caption{Limits on the effective DM-proton cross section for \ONR{9}. Proton coupling refers to the scenario where $\hat{c}_i^n=0.01\hat{c}_i^p$ while neutron coupling takes $\hat{c}_i^p=0.01\hat{c}_i^n$. The quark coupling case uses the ratios given in Table \ref{tab:coupling_ratios}.}
    \label{fig:O9_limits}
\end{figure*}

\begin{figure*}[htpb]
    \centering
    \includegraphics[width=\linewidth]{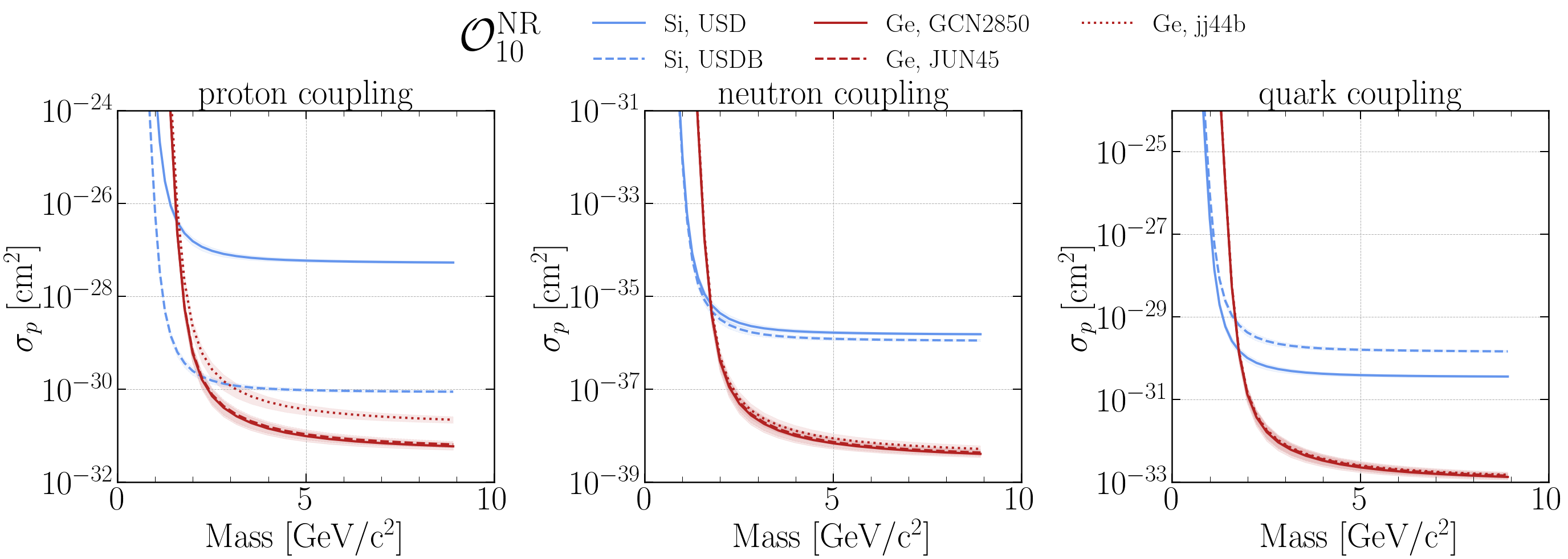}
    \captionsetup{justification=Justified}
    \caption{Limits on the effective DM-proton cross section for \ONR{10}. Proton coupling refers to the scenario where $\hat{c}_i^n=0.01\hat{c}_i^p$ while neutron coupling takes $\hat{c}_i^p=0.01\hat{c}_i^n$. The quark coupling case uses the ratios given in Table \ref{tab:coupling_ratios}.}
    \label{fig:O10_limits}
\end{figure*}

\begin{figure*}[htpb]
    \centering
    \includegraphics[width=\linewidth]{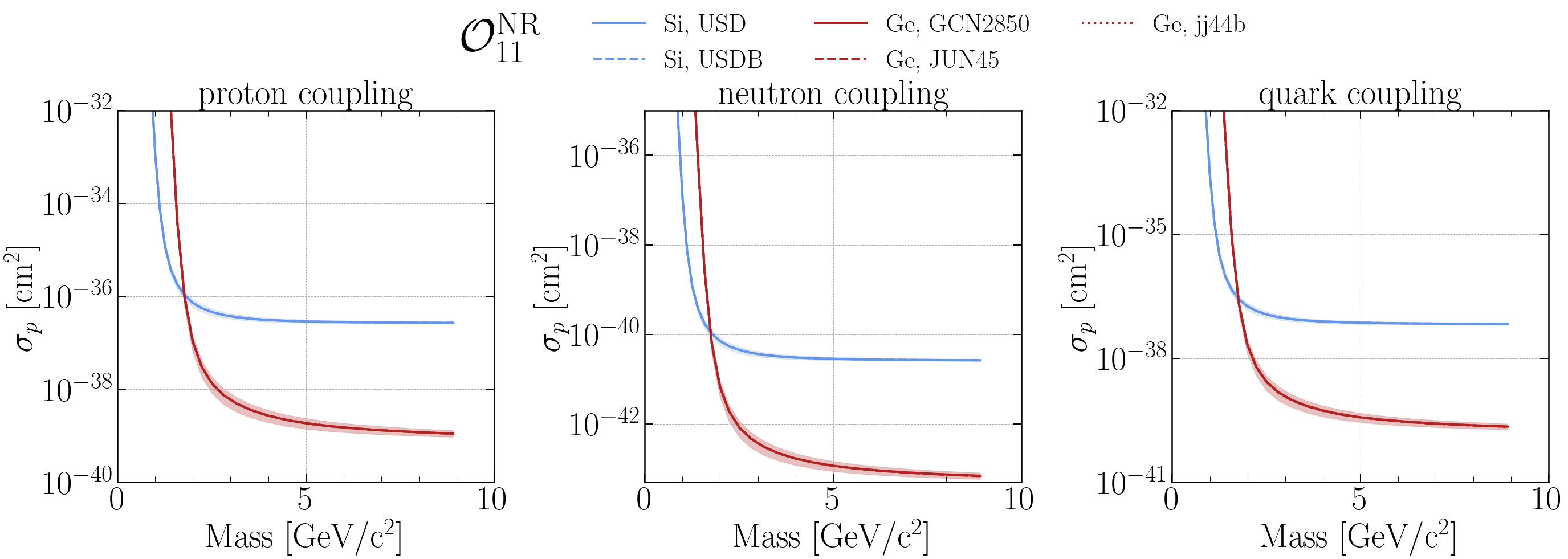}
    \captionsetup{justification=Justified}
    \caption{Limits on the effective DM-proton cross section for \ONR{11}. Proton coupling refers to the scenario where $\hat{c}_i^n=0.01\hat{c}_i^p$ while neutron coupling takes $\hat{c}_i^p=0.01\hat{c}_i^n$. The quark coupling case uses the ratios given in Table \ref{tab:coupling_ratios}.}
    \label{fig:O11_limits}
\end{figure*}

\begin{figure*}[htpb]
    \centering
    \includegraphics[width=\linewidth]{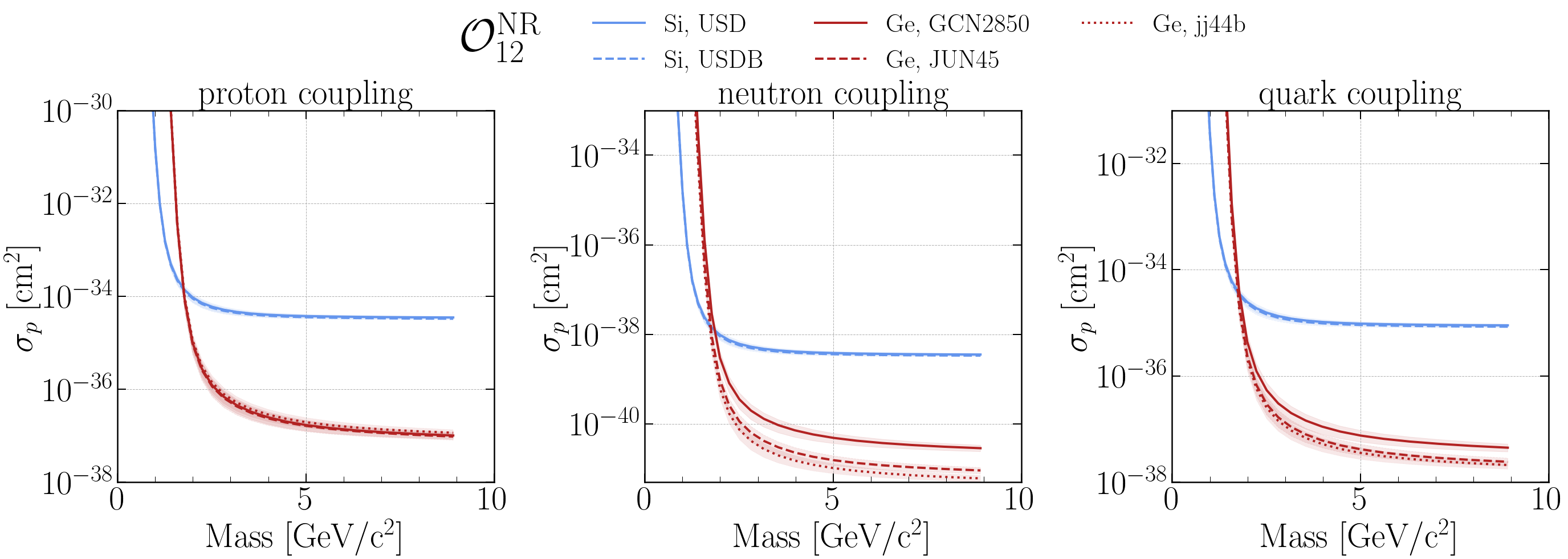}
    \captionsetup{justification=Justified}
    \caption{Limits on the effective DM-proton cross section for \ONR{12}. Proton coupling refers to the scenario where $\hat{c}_i^n=0.01\hat{c}_i^p$ while neutron coupling takes $\hat{c}_i^p=0.01\hat{c}_i^n$. The quark coupling case uses the ratios given in Table \ref{tab:coupling_ratios}.}
    \label{fig:O12_limits}
\end{figure*}

Using the optimum interval sensitivity method \cite{Yellin_2002,yellin2007extendOI}, we produce exclusion projections for Ge and Si targets under the assumptions given in Table \ref{tab:exp_params}. These are collected in Figs. \ref{fig:O1_limits}-\ref{fig:O12_limits}, with summaries given in Table \ref{tab:ratios_table}. This table gives the ratios of the exclusion limits for different nuclear structure factors. We note that there is almost no mass dependence on this ratio, suggesting that the rate under the different nuclear model assumptions are related by a multiplicative factor. We also include in this table which nuclear channels contribute to the various NR operators, and the relative integrated form factors (IFFs) for each nuclear structure model, where these IFF values are presented for $q_{\rm max}=50$ MeV. 

\begin{table*}[htpb]
\captionsetup{justification=Justified}
\caption{Ratio of nuclear IFFs and resulting sensitivity ratios, where the IFFs are calculated using $q_{\rm max}=50$ MeV, and $k=q/m_N$. Note that in general we would expect an increase in IFF to correspond to an increase in the observation rate, and thus a lower sensitivity.}
\label{tab:ratios_table}
\begin{tabular}{|c|ccc|ccc|}
\hline 
\multirow{2}{*}{\textbf{NR operator}} & \multicolumn{3}{c|}{\textbf{Integrated nuclear form factor ratio}} & \multicolumn{3}{c|}{\textbf{Sensitivity ratio}} \\
 & Nuclear operator & p & n & Proton coupling & Neutron coupling & Quark coupling \\ \hline 
\multicolumn{7}{|c|}{\textbf{Si: USDB/USD}} \\ \hline 
\ONR{1} & $M$ & 1.0 & 1.0 & 1.00 $\pm$ 0.24 & 1.02 $\pm$ 0.25 & 1.00 $\pm$ 0.24 \\ \hline
\multirow{2}{*}{\ONR{3}} & $k^2 \Sigma'$ & 69& 1.4 & \multirow{2}{*}{0.94 $\pm$ 0.18} & \multirow{2}{*}{0.94 $\pm$ 0.18} & \multirow{2}{*}{0.94 $\pm$ 0.18} \\
 & $k^4 \Phi''$ & 1.1 & 1.1 &  &  &  \\ \hline
\multirow{2}{*}{\ONR{4}} & $\Sigma'$ & 67& 1.4 & \multirow{2}{*}{(1.6 $\pm$ 0.4) $\times 10^{-3}$} & \multirow{2}{*}{0.74 $\pm$ 0.18} & \multirow{2}{*}{0.59 $\pm$ 0.14} \\
 & $\Sigma''$ & 53& 1.3 &  &  &  \\ \hline
\multirow{2}{*}{\ONR{5}} & $k^2 M$ & 1.0 & 1.0 & \multirow{2}{*}{0.99 $\pm$ 0.21} & \multirow{2}{*}{1.00 $\pm$ 0.21} & \multirow{2}{*}{1.00 $\pm$ 0.22} \\
 & $k^4 \Delta$ & 0.007& 0.91 &  &  &  \\ \hline
\ONR{6} & $k^4 \Sigma''$ & 49& 1.3 & (1.6 $\pm$ 0.3) $\times 10^{-3}$ & 0.74 $\pm$ 0.14 & 4.08 $\pm$ 0.8 \\ \hline
\ONR{7} & $\Sigma'$ & 67& 1.4 & (1.6 $\pm$ 0.4) $\times 10^{-3}$ & 0.73 $\pm$ 0.18 & 0.58 $\pm$ 0.15 \\ \hline
\multirow{2}{*}{\ONR{8}} & $M$ & 1.0 & 1.0 & \multirow{2}{*}{0.99 $\pm$ 0.25} & \multirow{2}{*}{1.00 $\pm$ 0.25} & \multirow{2}{*}{0.99 $\pm$ 0.24} \\
 & $k^2 \Delta$ & 0.007 & 0.91 &  &  &  \\ \hline
\ONR{9} & $k^2 \Sigma'$ & 69& 1.4 & (1.6 $\pm$ 0.3) $\times 10^{-3}$ & 0.73 $\pm$ 0.16 & 0.58 $\pm$ 0.12 \\ \hline
\ONR{10} & $k^2 \Sigma''$ & 51& 1.3 & (1.6 $\pm$ 0.3) $\times 10^{-3}$ & 0.73 $\pm$ 0.15 & 4.08 $\pm$ 0.87 \\ \hline
\ONR{11} & $k^2 M$ & 1.0 & 1.0 & 1.00 $\pm$ 0.21 & 1.00 $\pm$ 0.21 & 1.00 $\pm$ 0.21 \\ \hline
\multirow{4}{*}{\ONR{12}} & $\Sigma'$ & 67& 1.4 & \multirow{4}{*}{0.95 $\pm$ 0.20} & \multirow{4}{*}{0.94 $\pm$ 0.20} & \multirow{4}{*}{0.94 $\pm$ 0.20} \\
 & $\Sigma''$ & 53& 1.3 &  &  &  \\
 & $k^2 \Phi''$ & 1.1 & 1.1 &  &  &  \\
 & $k^2 \tilde{\Phi}'$ & - & - &  &  &  \\ \hline 
\multicolumn{7}{|c|}{\textbf{Ge: GCN2850/JUN45}} \\ \hline 
\ONR{1} & $M$ & 1.0 & 1.0 & 1.00 $\pm$ 0.50 & 0.99 $\pm$ 0.48 & 1.00 $\pm$ 0.49 \\ \hline
\multirow{2}{*}{\ONR{3}} & $k^2 \Sigma'$ & 1.3& 1.1 & \multirow{2}{*}{1.04 $\pm$ 0.45} & \multirow{2}{*}{3.2 $\pm$ 1.4} & \multirow{2}{*}{1.83 $\pm$ 0.82} \\
 & $k^4 \Phi''$ & 1.0 & 0.313 &  &  &  \\ \hline
\multirow{2}{*}{\ONR{4}} & $\Sigma'$ & 1.3& 1.1 & \multirow{2}{*}{0.89 $\pm$ 0.43} & \multirow{2}{*}{0.92 $\pm$ 0.44} & \multirow{2}{*}{0.93 $\pm$ 0.44} \\
 & $\Sigma''$ & 1.0 & 1.1 &  &  &  \\ \hline
\multirow{2}{*}{\ONR{5}} & $k^2 M$ & 1.0 & 1.0 & \multirow{2}{*}{1.00 $\pm$ 0.45} & \multirow{2}{*}{0.98 $\pm$ 0.45} & \multirow{2}{*}{0.97 $\pm$ 0.44} \\
 & $k^4 \Delta$ & 0.588 & 1.0 &  &  &  \\ \hline
\ONR{6} & $k^4 \Sigma''$ & 1.0 & 1.1 & 0.90 $\pm$ 0.40 & 0.94 $\pm$ 0.41 & 0.91 $\pm$ 0.40 \\ \hline
\ONR{7} & $\Sigma'$ & 1.3& 1.1 & 0.88 $\pm$ 0.43 & 0.95 $\pm$ 0.47 & 0.94 $\pm$ 0.45 \\ \hline
\multirow{2}{*}{\ONR{8}} & $M$ & 1.0 & 1.0 & \multirow{2}{*}{1.00 $\pm$ 0.49} & \multirow{2}{*}{0.99 $\pm$ 0.48} & \multirow{2}{*}{1.02 $\pm$ 0.49} \\
 & $k^2 \Delta$ & 0.588 & 1.0 &  &  &  \\ \hline
\ONR{9} & $k^2 \Sigma'$ & 1.3& 1.1 & 0.88 $\pm$ 0.40 & 0.93 $\pm$ 0.43 & 0.92 $\pm$ 0.41 \\ \hline
\ONR{10} & $k^2 \Sigma''$ & 1.0 & 1.1 & 0.91 $\pm$ 0.41 & 0.93 $\pm$ 0.43 & 0.92 $\pm$ 0.42 \\ \hline
\ONR{11} & $k^2 M$ & 1.0 & 1.0 & 0.99 $\pm$ 0.46 & 1.00 $\pm$ 0.47 & 1.01 $\pm$ 0.47 \\ \hline
\multirow{4}{*}{\ONR{12}} & $\Sigma'$ & 1.3& 1.1 & \multirow{4}{*}{1.07 $\pm$ 0.49} & \multirow{4}{*}{3.2 $\pm$ 1.5} & \multirow{4}{*}{1.83 $\pm$ 0.82} \\
 & $\Sigma''$ & 1.0 & 1.1 &  &  &  \\
 & $k^2 \Phi''$ & 1.0 & 0.313 &  &  &  \\
 & $k^2 \tilde{\Phi}'$ & 42& 5.5&  &  &  \\ \hline  
\multicolumn{7}{|c|}{\textbf{Ge: GCN2850/jjb44}} \\ \hline 
\ONR{1} & $M$ & 1.0 & 1.0 & 0.99 $\pm$ 0.47 & 1.01 $\pm$ 0.49 & 1.00 $\pm$ 0.49 \\ \hline
\multirow{2}{*}{\ONR{3}} & $k^2 \Sigma'$ & 12.74& 1.32 & \multirow{2}{*}{0.86 $\pm$ 0.37} & \multirow{2}{*}{4.8 $\pm$ 2.1} & \multirow{2}{*}{2.10 $\pm$ 0.92} \\
 & $k^4 \Phi''$ & 1.2 & 0.208&  &  &  \\ \hline
\multirow{2}{*}{\ONR{4}} & $\Sigma'$ & 13& 1.32 & \multirow{2}{*}{0.26 $\pm$ 0.13} & \multirow{2}{*}{0.80 $\pm$ 0.39} & \multirow{2}{*}{0.78 $\pm$ 0.38} \\
 & $\Sigma''$ & 14& 1.32 &  &  &  \\ \hline
\multirow{2}{*}{\ONR{5}} & $k^2 M$ & 1.0 & 1.0 & \multirow{2}{*}{1.0 $\pm$ 0.45} & \multirow{2}{*}{0.98 $\pm$ 0.45} & \multirow{2}{*}{0.99 $\pm$ 0.45} \\
 & $k^4 \Delta$ & 0.29& 1.0&  &  &  \\ \hline
\ONR{6} & $k^4 \Sigma''$ & 18& 1.32 & 0.27 $\pm$ 0.12 & 0.79 $\pm$ 0.35 & 0.89 $\pm$ 0.40 \\ \hline
\ONR{7} & $\Sigma'$ & 13& 1.32 & 0.26 $\pm$ 0.12 & 0.79 $\pm$ 0.39 & 0.78 $\pm$ 0.37 \\ \hline
\multirow{2}{*}{\ONR{8}} & $M$ & 1.0 & 1.0 & \multirow{2}{*}{0.99 $\pm$ 0.48} & \multirow{2}{*}{0.97 $\pm$ 0.47} & \multirow{2}{*}{1.02 $\pm$ 0.50} \\
 & $k^2 \Delta$ & 0.29& 1.0&  &  &  \\ \hline
\ONR{9} & $k^2 \Sigma'$ & 12.74& 1.32 & 0.26 $\pm$ 0.12 & 0.80 $\pm$ 0.37 & 0.75 $\pm$ 0.34 \\ \hline
\ONR{10} & $k^2 \Sigma''$ & 16& 1.32 & 0.26 $\pm$ 0.12 & 0.78 $\pm$ 0.36 & 0.88 $\pm$ 0.40 \\ \hline
\ONR{11} & $k^2 M$ & 1.0 & 1.0 & 1.00 $\pm$ 0.46 & 1.00 $\pm$ 0.47 & 0.99 $\pm$ 0.45 \\ \hline
\multirow{4}{*}{\ONR{12}} & $\Sigma'$ & 13& 1.32 & \multirow{4}{*}{0.87 $\pm$ 0.40} & \multirow{4}{*}{4.9 $\pm$ 0.20} & \multirow{4}{*}{2.14 $\pm$ 0.20} \\
 & $\Sigma''$ & 14& 1.32 &  &  &  \\
 & $k^2 \Phi''$ & 1.2 & 0.208&  &  &  \\
 & $k^2 \tilde{\Phi}'$ & 63& 77&  &  &  \\\hline
 \end{tabular}
\end{table*}

At first glance, it can be difficult to see clear trends in these results given the interplay between the relative weighting of the form factors, and the choice of coupling constants. Indeed, given form factors from two nuclear shell model interactions, it is entirely possible to `tune' coupling constants that give you the maximal possible difference between the models. Thus, it is difficult to be definitive about when and if the choice of a particular nuclear shell model will impact experimental results. Instead, we offer the following observations of trends, and provide a metric to estimate the difference that might be introduced without requiring a full detector modelling and sensitivity recalculation. 

First, where multiple nuclear response terms contribute to an interaction rate their relative weights are set entirely by the nuclear shell model. That is to say, the coupling coefficient scenarios employed here (proton-dominant, neutron-dominant, and quark-level) retain the (sub-)dominant nature of each nuclear channel in the $F_{i,j}^{(N,N')}$ terms at the experimental limit level. As such, the nuclear uncertainties present in the dominant nuclear responses at the IFF level are expected to be the most important in experimental analysis.
This can be seen most clearly by examining the results for \ONR{8}. For both Ge and Si (and in all nuclear shell models), the IFF $k^2\Delta$ values are 7-8 orders of magnitude smaller than that of the $M$ term, meaning that the $\Delta$ factor could differ by several orders of magnitude between nuclear shell models and the total rates (in combination with the $M$ form factor) remain unchanged. Examining the IFFs in Figs.~\ref{fig: silicon 50 MeV IFFs} and \ref{fig: germanium 50 MeV IFFs} and the possible combinations of form factors in Table \ref{tab:form_fact} it is clear that the major drivers of uncertainty in experimental limits will be from changes to the $M$, $\Phi''$, $\Sigma'$, and $\Sigma''$ terms. 
Theoretically, it could be possible to tune a set of coupling constants such that the resulting interaction model is only sensitive to a single nuclear operator that might change the most between different shell model calculations. For example, to look soley at the $\Delta$ operator one could focus on specific combinations of \ONR{1} and \ONR{8} that cancel out the $M$ dependence. However, any such model is likely to be highly contrived and not well motivated from the high energy regime, though it may be of interest when considering experimental results that are in tension with each other, as different nuclear shell model calculations could then help resolve this tension.

Second, the choice of a particular particle interaction model will dictate whether a nuclear shell model increases or decreases the sensitivity of an experiment. That is to say, none of the nuclear shell model calculations explored here perform `best' globally. For example, the GCN2850 shell model result sets the weakest limit for the \ONR{3} operator, but tends to set slightly stronger limits (though within statistical uncertainty) for other operators. This is to be expected -- looking at the IFF plots, the different nuclear responses do not change uniformly under the assumption of a different nuclear shell model, and so the limit will depend on exactly which nuclear response terms are involved in an interaction. The  interaction models with the largest difference between shell model calculations are those that correspond to the nuclear operators which have the largest variation in IFF values -- the purely-SD models for Si (particularly the proton channels), and the LSD $\Phi''$-dependent models for germanium (particularly the neutron channels). \\

Third, the relative weights of nucleon contributions to a given target are important. Although (as noted) it is possible to tune the relative neutron and proton contributions by changing the ratio of their coupling constants, targets will tend to have a `dominant' coupling mode (dictated by the type of unpaired nucleon) that drives the limits. For Si and Ge, this is the neutron channel, hence limits associated with neutron-dominant coefficients are consistently stronger than the corresponding proton-dominant or quark level counterparts. Typically the quark level limits are either of a similar order as the proton coupling limits, or are at an order of magnitude that is in between the proton and neutron limits. The nuclear  uncertainties that propagate from the form factors to the limits can depend on the ratio of the proton-neutron coupling constants employed. Thus, changes to the neutron form factors will generally have a larger impact on limits than changes to the proton.
This can be seen by comparing the results for the quark level couplings vs. the proton and neutron dominant couplings, where for both Si and Ge, the quark level couplings tend to follow the same trends when switching between shell model calculations as the neutron couplings rather than the proton. The only exception to this is the \ONR{6} and \ONR{10} operators for Si, where the limit on the quark couplings gets weaker while both the neutron and proton limits are stronger. The reason for this is that these models have a negative proton-neutron coupling ratio, so the enhancement of both the neutron and proton nuclear factors for the USDB model results in an overall reduction of rate.

Finally, the impact of changes to the nuclear shell model seems to be independent of mass, with the resulting limit curves being proportional to each other. Part of this may be due to the use of the optimum interval method and a flat background, which may `wash out' any spectral changes in the rates at different masses. Examining plots of the nuclear response terms as a function of $q$ does suggest that there is little change in the spectral features for different shell models, but a more robust statistical approach for limit setting would be desirable to more fully understand this.

To provide a simpler way to assess the impact of assuming a new shell model calculation on experimental sensitivity, we can recognise that the total number of events in a detector $N_d$ will be proportional to the integral of the differential cross section:
\begin{equation}
    N_d \propto\sigma_{\chi}\int \sum_{i,j} \sum_{N,N'=p,n}  c^N_{i,{\rm NR}} c^{N'}_{j,{\rm NR}} F_{ij}^{(N,N')} (v^2,q^2) dq.
\end{equation}
We can assume we need some minimal $N_d$ in a detector to be sensitive to a model, and for any nuclear shell model we find the excluded cross section is given by

\begin{equation}
    \frac{1}{\sigma_{\chi}} \propto \int \sum_{i,j} \sum_{N,N'=p,n}  c^N_{i,{\rm NR}} c^{N'}_{j,{\rm NR}} F_{ij}^{(N,N')} (v^2,q^2).
\end{equation}

Integrating the cross section in Eq.~\ref{full_cross} using the same integral upper limit $q_{\rm max}$ as the IFFs, the nuclear uncertainties of both quantities can be linked give a rough indication of the degree of propagation of these uncertainties from the nuclear form factors to DM observables. For each target isotope $T$ and at each $m_\chi$ value, the ratio of the integrated cross section between two different shell model interactions, designated by $B_1$ and $B_2$, we can use
\begin{equation}
    \int_0^{q_{\rm max}} \frac{q dq}{2}  F^{(N,N')}_{ij}(v,q,B) \equiv {\rm IFF}_{i,j}^{(N,N')} (v,q_{\rm max},B),
\end{equation}
to write

\begin{equation}
\resizebox{\linewidth}{!}{%
$\begin{split}
     & \frac{\sigma_{\chi} (m_\chi, B_1)}{\sigma_{\chi} (m_\chi, B_2)}  \sim  \frac{\sum_{N, N'=n,p} \hat{c}_{i,{\rm NR}}^{N}\hat{c}_{i,{\rm NR}}^{N'} {\rm IFF}_{i,i}^{(N,N')} (v,q_{\rm max},B_2)}{\sum_{N, N'=n,p} \hat{c}_{i,{\rm NR}}^{N}\hat{c}_{i,{\rm NR}}^{N'} {\rm IFF}_{i,i}^{(N,N')} (v,q_{\rm max},B_1)}.
\end{split}$
}
\end{equation}

We denote the above IFF ratio 
\begin{equation}\label{eq: IFF ratio uncertainty}
\begin{split}
     & {\rm IFF}_{i,B_2/B_1} (\hat{c}_{i,{\rm NR}}^{p},\hat{c}_{i,{\rm NR}}^{n}) \\
     &\equiv \frac{\sum_{N, N'=n,p} \hat{c}_{i,{\rm NR}}^{N}\hat{c}_{i,{\rm NR}}^{N'} {\rm IFF}_{i,i}^{(N,N')} (v,q_{\rm max},B_2)}{\sum_{N, N'=n,p} \hat{c}_{i,{\rm NR}}^{N}\hat{c}_{i,{\rm NR}}^{N'} {\rm IFF}_{i,i}^{(N,N')} (v,q_{\rm max},B_1)},
\end{split}
\end{equation}

\noindent and use this as a metric to evaluate the degree to which the IFF nuclear uncertainties are magnified in the DM exclusion limits. This highly depends on the form factor uncertainties, as well as the proton-to-neutron coefficient ratio $\hat{c}_{i,{\rm NR}}^{p}/\hat{c}_{i,{\rm NR}}^{n}$. Before calculating the DM exclusions for different nuclear models and detailed detector response terms, this IFF ratio can be used as a first measure of the potential impact of nuclear uncertainties on the observables. Since ${\rm IFF}_{i,B_2/B_1} (\hat{c}_{i,{\rm NR}}^{p},\hat{c}_{i,{\rm NR}}^{n})$ applies for each isotope separately, it can also be used to gauge which isotopes of silicon and germanium contribute the most to the observable nuclear uncertainties for the various operators $\mathcal{O}_i^{\rm NR}$. Table~\ref{tab:IFF_ratios_coefficients} showcases this IFF ratio for the three coupling coefficient scenarios employed here, for all silicon and germanium isotopes. This summary can be employed to identify where the largest uncertainty propagation may occur between the nuclear form factors and DM observables.

\begin{table*}[htpb]
\captionsetup{justification=Justified}
\caption{The IFF ratio ${\rm IFF}_{i,B_2/B_1} (\hat{c}_{i,{\rm NR}}^{p},\hat{c}_{i,{\rm NR}}^{n})$ for silicon and germanium isotopes, for three different coefficient scenarios. We have employed: $\hat{c}_i^p=1$ \&  $\hat{c}_i^n=100$ (denoted by `n' in table);  $\hat{c}_i^p=100$ \&  $\hat{c}_i^n=1$ (denoted by `p' in table); and $\hat{c}_i^p/\hat{c}_i^n$ obtained from quark-level operators (denoted by `q' in table). Here, $m_\chi =5$ GeV.}
\label{tab:IFF_ratios_coefficients}
\begin{tabular}{|c|c|ccccccccccc|}
\hline 
\multicolumn{2}{|c}{} &  \boldsymbol{\ONR{1}} & \boldsymbol{\ONR{3}} & \boldsymbol{\ONR{4}} & \boldsymbol{\ONR{5}} & \boldsymbol{\ONR{6}} & \boldsymbol{\ONR{7}} &  \boldsymbol{\ONR{8}} & \boldsymbol{\ONR{9}} & \boldsymbol{\ONR{10}} & \boldsymbol{\ONR{11}} & \boldsymbol{\ONR{12}} \\ 
\hline
\multicolumn{13}{|c|}{\textbf{USDB/USD}} \\ 
\hline
\multirow{3}{*}{\bf{$^{28}$Si}} &   p & 1.0 & 1.06 & - & 1.0 & - & - & 1.0 & - & - & 1.0 & 1.06 \\ 
&  n & 1.0 & 1.06 & - & 1.0 & - & - & 1.0 & - & - & 1.0 & 1.06 \\
&  q & 1.0 & 1.06 & - & 1.0 & - & - & 1.0 & - & - & 1.0 & 1.06 \\ \hline
\multirow{3}{*}{\bf{$^{29}$Si}} &   p & 1.0 & 1.06 & 621 & 1.0 & 445 & 745 & 1.0 & 745  & 462 & 1.0 & 1.06 \\ 
&  n & 1.0 & 1.08 & 1.36 & 1.0 & 1.34 & 1.38 & 1.0 & 1.38 & 1.34 & 1.0  & 1.09\\
& q & 1.0 & 1.07 & 1.71 & 1.0 & 0.28 & 1.73 & 1.0 & 1.73 & 0.27 & 1.0 & 1.08 \\ \hline
\multirow{3}{*}{\bf{$^{30}$Si}} &  p & 1.0 & 1.06 & - & 1.0 & - & - & 1.0 & - & - & 1.0 & 1.06  \\ 
&  n & 1.0 & 1.08 & - & 1.0 & - & - & 1.0 & - & - & 1.0 & 1.08 \\
&  q & 1.0 & 1.07 & - & 1.0 & - & - & 1.0 & - & - & 1.0 & 1.07 \\ \hline \hline
 \multicolumn{13}{|c|}{\textbf{GCN2850/JUN45}} \\ 
\hline
\multirow{3}{*}{\bf{$^{70}$Ge}} &   p & 1.0 & 0.633 & - & 1.0 & - & - & 1.0 & - & - & 1.0  & 0.633 \\ 
&  n & 1.0 & 0.639 & - & 1.0 & - & - & 1.0 & - & - & 1.0 & 0.639  \\
&  q & 1.0 & 0.636 & - & 1.0 & - & - & 1.0 & - & - & 1.0 & 0.636 \\ \hline
\multirow{3}{*}{\bf{$^{72}$Ge}} &   p & 1.0 & 1.316 & - & 1.0 & - & - & 1.0 & - & - & 1.0 & 1.316 \\ 
&  n & 1.0 & 0.022 & - & 1.0 & - & - & 1.0 & - & - & 1.0 & 0.022 \\
& q & 1.0 & 0.471 & - & 1.0 & - & - & 1.0 & - & - & 1.0 & 0.471 \\ \hline
\multirow{3}{*}{\bf{$^{73}$Ge}} &  p & 1.0 & 1.09 & 1.13 & 1.0 & 0.99 & 1.21 & 1.0 & 1.21 & 1.01 & 1.0 & 1.09 \\ 
&  n & 1.0 & 0.15 & 1.07 & 1.0 & 1.07 & 1.07 & 1.0 & 1.07 & 1.07 &  1.0 & 0.17 \\
&  q & 1.0 & 0.51 & 1.08 & 1.0 & 1.09 & 1.08 & 1.0 & 1.08 & 1.09 & 1.0 & 0.51 \\ \hline
\multirow{3}{*}{\bf{$^{74}$Ge}} &  p & 1.0 & 0.85& - & 1.0 & - & - & 1.0 & - & - & 1.0 & 0.85 \\ 
&  n & 1.0 & 0.32 & - & 1.0 & - & - & 1.0 & - & - & 1.0 & 0.32  \\
&  q & 1.0 & 0.53 & - & 1.0 & - & - & 1.0 & - & - & 1.0 & 0.53 \\ \hline
\multirow{3}{*}{\bf{$^{76}$Ge}} &  p & 1.0 & 0.82 & - & 1.0 & - & - & 1.0 & - & - & 1.0 & 0.82 \\ 
&  n & 1.0 & 0.60 & - & 1.0 & - & - & 1.0 & - & - & 1.0 & 0.60 \\
&  q & 1.0 & 0.69 & - & 1.0 & - & - & 1.0 & - & - & 1.0 & 0.69 \\ \hline
\hline
 \multicolumn{13}{|c|}{\textbf{GCN2850/jj44b}} \\ 
\hline 
\multirow{3}{*}{\bf{$^{70}$Ge}} &   p & 1.0 & 0.84 & - & 1.0 & - & - & 1.0 & - & - & 1.0 & 0.84 \\ 
&  n & 1.0 & 0.34 & - & 1.0 & - & - & 1.0 & - & - & 1.0 & 0.34 \\
&  q & 1.0 & 0.55 & - & 1.0 & - & - & 1.0 & - & - & 1.0 & 0.55 \\ \hline
\multirow{3}{*}{\bf{$^{72}$Ge}} &   p & 1.0 & 1.684 & - & 1.0 & - & - & 1.0 & - & - & 1.0 & 1.683 \\ 
&  n & 1.0 & 0.013 & - & 1.0 & - & - & 1.0 & - & -& 1.0 & 0.013 \\
& q & 1.0 & 0.400 & - & 1.0 & - & - & 1.0 & - & -& 1.0 & 0.399 \\ \hline
\multirow{3}{*}{\bf{$^{73}$Ge}} &  p & 1.0 & 1.356 & 3.94 & 1.0 & 3.84 & 4.10 & 1.0 & 4.10 & 3.83 & 1.0 & 1.36 \\ 
&  n & 1.0 & 0.089 & 1.28 & 1.0 & 1.30 & 1.27 & 1.0 & 1.27 & 1.29 & 1.0 & 0.10 \\
&  q & 1.0 & 0.419 & 1.31 & 1.0 & 1.16 & 1.30 & 1.0 & 1.30 & 1.16 & 1.0 & 0.42 \\ \hline
\multirow{3}{*}{\bf{$^{74}$Ge}} &  p & 1.0 & 0.95 & - & 1.0 & - & - & 1.0 & - & -& 1.0 & 0.95 \\ 
&  n & 1.0 & 0.23 & - & 1.0 & - & - & 1.0 & - & - & 1.0 & 0.23 \\
&  q & 1.0 & 0.46 & - & 1.0 & - & - & 1.0 & - & - & 1.0 & 0.46  \\ \hline
\multirow{3}{*}{\bf{$^{76}$Ge}} &  p & 1.0 & 0.88 & - & 1.0 & - & - & 1.0 & - & - & 1.0 & 0.88 \\ 
&  n & 1.0 & 0.49 & - & 1.0 & - & - & 1.0 & - & - & 1.0 & 0.49 \\
&  q & 1.0 & 0.62 & - & 1.0 & - & - & 1.0 & - & - & 1.0 & 0.62 \\ \hline
\end{tabular}
\end{table*}

\section{Conclusions}
\label{sec:conc}

In this work we have quantified silicon and germanium shell model nuclear uncertainties present in eleven operators within the framework of non-relativistic effective field theory (NREFT), relevant for WIMP-nucleus elastic scattering theories and experiments. This approach considers a more complete framework for nuclear modelling, with the inclusion of additional nuclear operators compared to the standard SI and SD channels, namely operators which depend on the orbital angular momentum $L$ (LD) as well as on both $s$ (spin) and $L$ (LSD). 

The $^{28,29,30}$Si and $^{70,72,73,74,76}$Ge nuclear form factors were evaluated through large-scale shell model calculations, employing several shell model interactions -- USD and USDB for silicon, and GCN2850, JUN45 and jj44b for germanium. Nuclear uncertainties were quantified through an integrated form factor (IFF) value for each nuclear operator. The degree of propagation of the nuclear uncertainties to WIMP scattering observables was analysed through exclusion limits, produced assuming SuperCDMS-like detectors using the optimum interval limit setting method.

The impact of these nuclear shell model interactions on sensitivity were presented under different coupling assumptions for protons and neutrons to understand how this would enhance or suppress specific nuclear response channels, which themselves behave differently depending on the shell model calculation. We find that the interplay between these is non-trivial, but that the IFF ratio ${\rm IFF}_{i,B_2/B_1} (\hat{c}_{i,{\rm NR}}^{p},\hat{c}_{i,{\rm NR}}^{n})$ can be used to estimate the change in sensitivity one would expect for a given particle interaction model and nuclear shell model calculation, and indicate whether the uncertainties will be magnified or suppressed in the exclusion limit calculation.

These results highlight the importance of accounting for nuclear modelling uncertainties in dark matter direct detection searches, through a more complete nuclear response framework. Shell model interaction choice in silicon and germanium calculations is associated with non-negligible nuclear uncertainties for a range of the DM nuclear responses. This may greatly impact DM experimental analysis and predictions, particularly those that change the nuclear response operators $M$, $\Phi''$, $\Sigma'$, and $\Sigma''$. Further nuclear uncertainty quantification studies should be performed for nuclear models other than the shell model, to better ascertain the impact of nuclear modelling on DM experimental observables. 

\section{Acknowledgements}

R.~AK was supported by the Australian Government through the Australian Research Council Centre of Excellence for Dark Matter Particle Physics (CDM, CE200100008). 
MJZ would like to acknowledge the support of the Natural Sciences and Engineering Research Council of Canada and the Arthur B. McDonald Canadian Astroparticle Physics Research Institute

\newpage
\def\bibsection{\section*{\refname}}
 
\bibliography{bib}

\appendix

\section{Unitless vs. unitful form factors}
\label{sec:ff_units}
There exist two sets of non-relativistic operators that can be used for calculating DM interaction rates: `unitful' non-relativistic operators (used in Refs.~\cite{Fitzpatrick:2012ix,Cirelli_2013}) and `unitless' non-relativistic operators (used in Ref.~\cite{Anand:2013yka}). These are given in Table~\ref{tab:nr_operators_comp}. We give them these labels because the resulting expressions are form factors that contain explicit units (in the case of the `unitful` expressions) or do not (in the case of the `unitless').

\begin{table}[!h]
\captionsetup{justification=Justified}
\caption{The non-relativistic operators assumed in the unitful  \cite{Fitzpatrick:2012ix} vs. unitless \cite{Anand:2013yka} expressions. Note that the only difference is the transformation $\vec{q}\rightarrow \vec{q}/m_N$.}
\label{tab:nr_operators_comp}
\begin{tabular}{@{}cll@{}}
\toprule
\textbf{Operator} & \textbf{Unitful expression} & \textbf{Unitless expression} \\ \midrule
\ONR{1} & \bb 1 & \bb 1 \\
\ONR{3} &  $i \, \vec{s}_N \cdot (\vec{q} \times \vec{v}^\perp)$ & $i \, \vec{s}_N \cdot \left(\frac{\vec{q}}{m_N} \times \vec{v}^\perp \right)$ \\
\ONR{4} & $\vec{s}_\chi \cdot \vec{s}_N$ & $\vec{s}_\chi \cdot \vec{s}_N$ \\
\ONR{5} &  $i \, \vec{s}_\chi \cdot (\vec{q} \times \vec{v}^\perp)$ & $i \, \vec{s}_\chi \cdot \left(\frac{\vec{q}}{m_N} \times \vec{v}^\perp \right)$ \\
\ONR{6} & $(\vec{s}_\chi \cdot \vec{q}) (\vec{s}_N \cdot \vec{q})$ & $\left(\vec{s}_\chi \cdot \frac{\vec{q}}{m_N}\right) \left(\vec{s}_N \cdot \frac{\vec{q}}{m_N}\right)$ \\
\ONR{7} & $\vec{s}_N \cdot \vec{v}^\perp$ & $\vec{s}_N \cdot \vec{v}^\perp$ \\
\ONR{8} & $\vec{s}_\chi \cdot \vec{v}^\perp$ & $\vec{s}_\chi \cdot \vec{v}^\perp$ \\
\ONR{9} & $i \, \vec{s}_\chi \cdot (\vec{s}_N \times \vec{q})$ & $i \, \vec{s}_\chi \cdot \left(\vec{s}_N \times \frac{\vec{q}}{m_N}\right)$ \\
\ONR{10} & $i \, \vec{s}_N \cdot \vec{q}$ & $i \, \vec{s}_N \cdot \frac{\vec{q}}{m_N}$ \\
\ONR{11} & $i \, \vec{s}_\chi \cdot \vec{q}$ & $i \, \vec{s}_\chi \cdot \frac{\vec{q}}{m_N}$ \\
\ONR{12} & $\vec{v}^\perp \cdot (\vec{s}_\chi \times \vec{s}_N)$ & $\vec{v}^\perp \cdot \left(\vec{s}_\chi \times \vec{s}_N \right)$ \\ \bottomrule
\end{tabular}
\end{table}

Although the difference between the two amounts to a transformation of $\vec{q}\rightarrow \vec{q}/m_N$, because not all operators have the same $q$ dependence there is no simple transformation that can be applied to all operators identically. Compared to the unitless form factors in Table~\ref{tab:form_fact}, the unitful counterparts are obtained through the transformation
\begin{equation}
\label{eq:cir_ff}
    \begin{aligned}
        & F_{3,3}^{(N,N')} \rightarrow  m_N^2 F_{3,3}^{(N,N')},
        & \hspace{3mm}
        & F_{5,5}^{(N,N')} \rightarrow m_N^2 F_{5,5}^{(N,N')},  \\ 
        & F_{6,6}^{(N,N')} \rightarrow m_N^4 F_{6,6}^{(N,N')},
        & \hspace{3mm}
        & F_{9,9}^{(N,N')} \rightarrow m_N^2 F_{9,9}^{(N,N')}, \\
        & F_{10,10}^{(N,N)'} \rightarrow m_N^2 F_{10,10}^{(N,N)'},
        & \hspace{3mm}
        & F_{11,11}^{(N,N')} \rightarrow m_N^2 F_{11,11}^{(N,N')}, \\
        & F_{1,3}^{(N,N')} \rightarrow m_N F_{1,3}^{(N,N')} ,
        & \hspace{3mm}
        & F_{4,5}^{(N,N')} \rightarrow m_N F_{4,5}^{(N,N')}, \\
        & F_{4,6}^{(N,N')} \rightarrow m_N^2 F_{4,6}^{(N,N')}, 
        & \hspace{3mm}
        & F_{9,8}^{(N,N')} \rightarrow m_N F_{9,8}^{(N,N')}, \\
        &  F_{11,12}^{(N,N')} \rightarrow  m_N F_{11,12}^{(N,N')},\\
    \end{aligned}
\end{equation}

The proportionality constant, $A_{\rm prop}$, for the two methods are derived from Refs. \cite{Fitzpatrick:2012ix,Anand:2013yka,Cirelli_2013}:
\begin{equation}
\label{eq:xsec_const}
\begin{split}
    A_{\rm unitful} &= \frac{m_T}{32 \pi m_\chi^2 m_N^2v^2}, \\
    A_{\rm unitless} &= \frac{m_T}{2\pi v^2},
\end{split}
\end{equation}
which can be used to calculate the set of coupling constants under one assumption, given the values under the other.

\section{Quark level couplings}
\label{sec:quark_couplings}

The full derivation of quark and gluon level couplings can be found in Ref.~\cite{Cirelli_2013}. Here we simply quote the resulting couplings, noting that $c^q_k$ and $c^g_k$ are real {\it dimensionful} coefficients: $c^q_k$ will have dimensions of [mass]$^{-2}$ and $c^g_k$ of [mass]$^{-3}$. These are inherited from factors of $\Lambda$ (units of eV) which is the scale of the new physics. Typically, for the quark scalar couplings we assume:
\begin{equation}
     c_i^q = \left\{\begin{matrix}
     m_q/(\Lambda^2)& i\leq 4  \\
     1/\Lambda^3& i>4 \\
    \end{matrix}\right.
\end{equation}
These then give relativistic nucleon couplings of
\begin{equation}
\begin{split}
c^N_{1, 2} &= \sum_{q = u, d, s} c^q_{1, 2} \frac{m_N}{m_q} f_{Tq}^{(N)} + \frac{2}{27} f_{TG}^{(N)} \\
& \times \left( \sum_{q = c, b, t} c^q_{1, 2} \frac{m_N}{m_q} - c^g_{1, 2} m_N \right) \ , 
\\
c^N_{3, 4} &= \sum_{q = u, d, s} \frac{m_N}{m_q} \left[ (c^q_{3, 4} - C_{3, 4}) + c^g_{3, 4} \bar{m} \right] \Delta_q^{(N)} \ ,
\\
c^p_{5, 6} &= 2 \, c^u_{5, 6} + c^d_{5, 6} \ ,
\quad
c^n_{5, 6} = c^u_{5, 6} + 2 \, c^d_{5, 6} \ ,
\\
c^N_{7, 8} &= \sum_q c^q_{7, 8} \, \Delta_q^{(N)} \ ,
\\
c^N_{9, 10} &= \sum_q c^q_{9, 10} \, \delta_q^{(N)} \ .
\end{split}
\end{equation}
where $C_{3, 4} \equiv \sum_q c^q_{3, 4} \, \bar{m} / m_q$ with $\bar{m} \equiv (1 / m_u + 1 / m_d + 1 / m_s)^{-1}$. The scalar and axial charges ($f_{Tq}$, $\Delta_q^{(N)}$) are provided in Table~\ref{nuclearities}, where the scalar charges $f_{TG}^{(N)}$ associated with the gluon contribution of the matrix element are expressed in terms of the light quarks through $f_{TG}^{(N)} \equiv 1- \sum\limits_{q=u,d,s} f_{Tq}^{(N)}$. The light quarks satisfy the following relations: $\Delta_u^{(p)} = \Delta_d^{(n)}$, $\Delta_d^{(p)} = \Delta_u^{(n)}$, $\Delta_s^{(p)} = \Delta_s^{(n)}$. For the heavy counterparts these are taken to be negligible \cite{Polyakov:1998rb,Cirelli_2013}. The values of the tensor charges are taken to be $\delta_u^{(p)} = 0.84$, $\delta_d^{(p)} = -0.23$, and $\delta_s^{(p)} = -0.05$ \cite{Belanger:2013oya, Belanger:2008sj}. The proton tensor charges are related to the neutron ones in the same way as the axial charges $\Delta_q^{(N)}$ above.

\begin{table}[htpb]
\centering
\captionsetup{justification=Justified}
\caption{\label{nuclearities} Values for the scalar and axial charges, taken from Ref.~\cite{Gondolo:2004sc}.}
\begin{tabular}{c|c|c|c|c|c|c|c}
$f_{Tu}^{(p)}$ & $f_{Tu}^{(n)}$ & $f_{Td}^{(p)}$ & $f_{Td}^{(n)}$ & $f_{Ts}^{(N)}$ & $\Delta_u^{(p)}$ & $\Delta_d^{(p)}$ & $\Delta_s^{(p)}$ \\
\hline
 $0.023$ & $0.019$ & $0.034$ & $0.041$ & $0.14$ & $0.77$ & $-0.40$ & $-0.12$ \\
\end{tabular}
\end{table}

Once defined, these relativistic couplings can be used to derive the non-relativistic coefficients used with the form factors in Eq.~\ref{eq:diff_xsec}.
Given that the original quark level definitions were given under the assumption of unitful operators, it is convenient to first define these following Ref.~\cite{Cirelli_2013}. To avoid confusion with the unitless couplings used in the rest of this text, we use bold font to identify these unitful couplings:
\begin{equation}
    \begin{split}
        \textbf{c}_{1, \text{NR}}^N & = 4 m_\chi m_N c_1^N + 4 m_\chi m_N c_5^N,\\ 
        \textbf{c}_{4, \text{NR}}^N & = -16 m_\chi m_N c_8^N + 32 m_\chi m_N c_9^N,\\
        \textbf{c}_{6, \text{NR}}^N & = 4 c_4^N,\\ 
        \textbf{c}_{7, \text{NR}}^N & = -8 m_\chi m_N c_7^N,\\
        \textbf{c}_{8, \text{NR}}^N & = 8 m_\chi m_N c_6^N,\\
        \textbf{c}_{9, \text{NR}}^N & = 8 m_\chi c_6^N + 8 m_N c_7^N,\\
        \textbf{c}_{10, \text{NR}}^N & = 4 m_\chi c_3^N - 8 m_N c_{10}^N,\\
        \textbf{c}_{11, \text{NR}}^N & = -4 m_N c_2^N + 8 m_\chi c_{10}^N,\\
        \textbf{c}_{12, \text{NR}}^N & = -32 m_\chi m_N c_{10}^N.\\
    \end{split}
    \label{nr_coup}
\end{equation}
Note that $\textbf{c}_{3, \text{NR}}^N$ and $\textbf{c}_{5, \text{NR}}^N$ do not have any mapping to the higher energy EFT operators employed here. This is equivalent to saying there is no theoretically motivated interaction model that can produce an experimental rate driven by those non relativistic operators.

To find the equivalent relationship for the unitless construction, we can assume that regardless of which construction is used, the same expression for the differential cross section should be derived. We can then take advantage of Eq.~\ref{eq:xsec_const} to relate the unitless form factor and couplings (in regular text) to the unitful expressions (in bold text):
\begin{equation}
    \left(c_{i,{\rm NR}}^N\right)^2 = \frac{1}{16 m^2_{\chi} m^2_N}\left(\textbf{c}_{i,{\rm NR}}^N\right)^2\frac{\bold{F}^{N,N}_{i,i}}{F^{N,N}_{i,i}}.
\end{equation}
This then allows us to derive expressions for the coupling constants for the unitless form factors from the same set of quark couplings as for the unitful:
\begin{equation}
    \begin{split}
        c_{1, \text{NR}}^N & = c_1^N + c_5^N,\\ 
        c_{4, \text{NR}}^N & = -4 c_8^N + 8 c_9^N,\\
        c_{6, \text{NR}}^N & = \frac{m_N}{m_{\chi}} c_4^N,\\ 
        c_{7, \text{NR}}^N & = -2 c_7^N,\\
        c_{8, \text{NR}}^N & = 2 c_6^N,\\
        c_{9, \text{NR}}^N & = 2 c_6^N + 2 \frac{m_N}{m_{\chi}} c_7^N,\\
        c_{10, \text{NR}}^N & = c_3^N - 2 \frac{m_N}{m_{\chi}} c_{10}^N,\\
        c_{11, \text{NR}}^N & = -\frac{m_N}{m_{\chi}} c_2^N + 2 c_{10}^N,\\
        c_{12, \text{NR}}^N & = 8c^N_{10}.\\
    \end{split}
    \label{cross_to_coup}
\end{equation}

We note that one must be careful when employing expressions from both Refs.~\cite{Fitzpatrick:2012ix,Anand:2013yka} and Ref.~\cite{Cirelli_2013}, as one uses an inverted definition of the momentum transfer $\vec{q}$ compared to the other. In this case, an additional minus sign must be accounted for in some nuclear operators and form factors.

\begin{figure*}[htpb]
    \centering
    \includegraphics[width=0.355\linewidth]{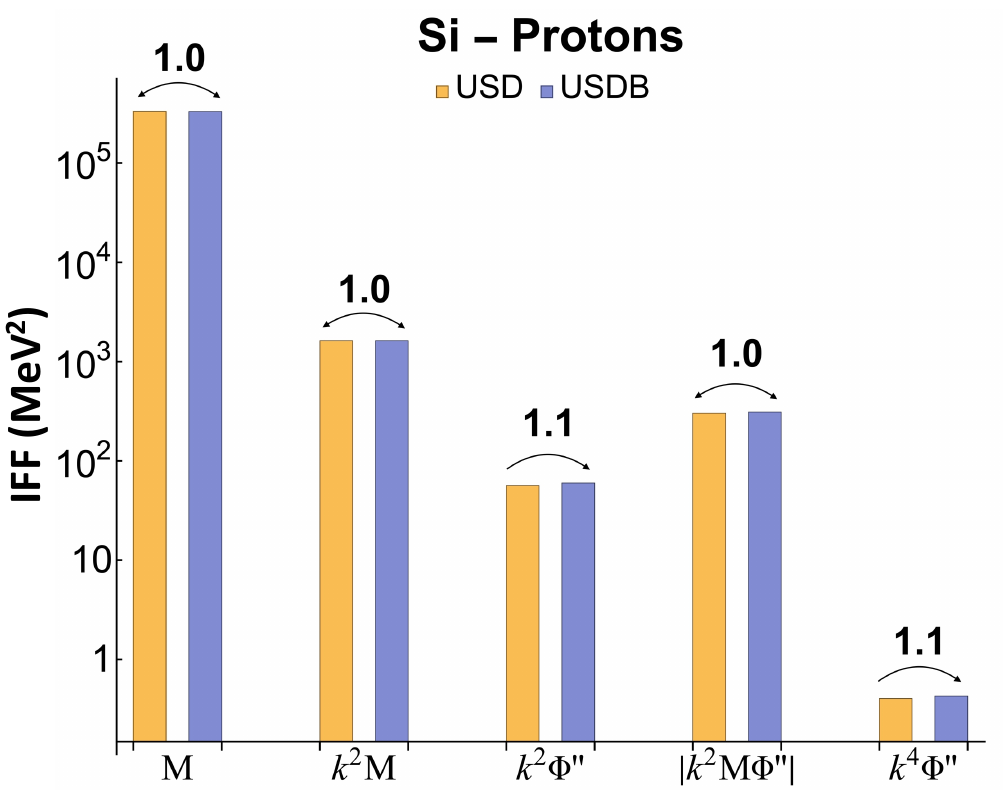}~
    \includegraphics[width=0.355\linewidth]{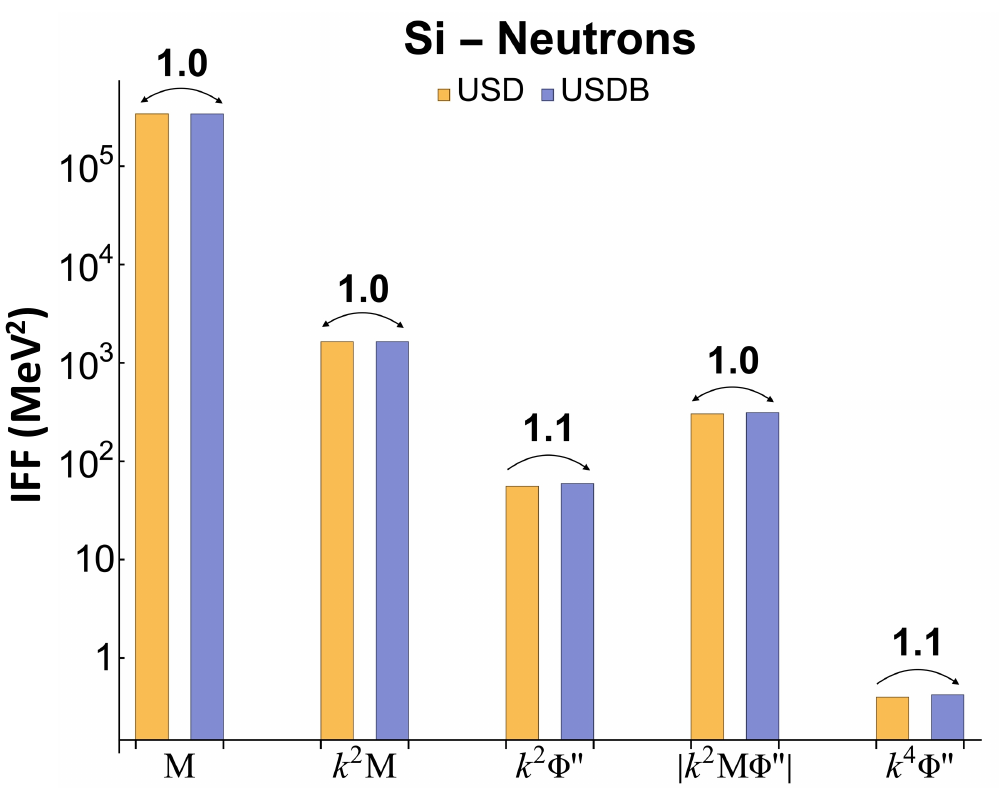}\\
    \includegraphics[width=0.355\linewidth]{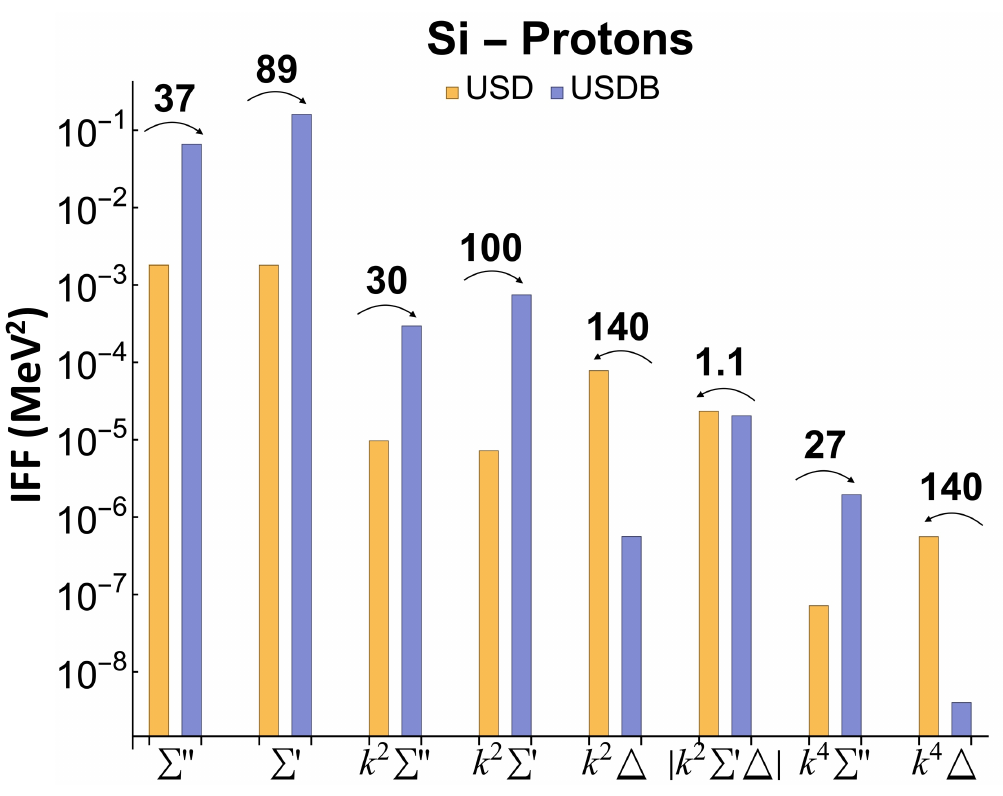}~
    \includegraphics[width=0.355\linewidth]{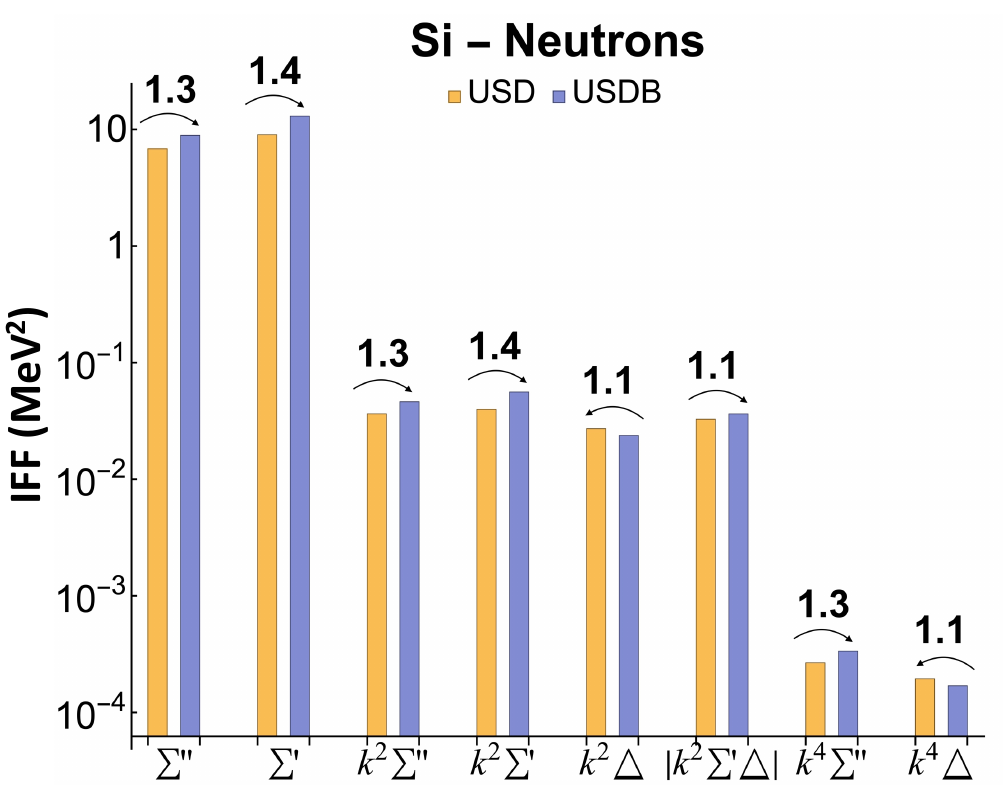}
    \captionsetup{justification=Justified}
    \caption{Silicon proton (left) and neutron (right) IFF values for an integral with $q_{\rm max}= 100$ MeV. Top panel shows values related to the nuclear responses $M$, $\Phi''$, whilst bottom panel shows those related to $\Sigma''$, $\Sigma'$, $\Delta$.}
    \label{fig: silicon 100 MeV IFFs}
\end{figure*}

\section{IFF Plots For $q_{\rm max}=100$ MeV}\label{appendix: IFFs 100 MeV}

\begin{figure*}[htpb]
    \centering
    \includegraphics[width=0.355\linewidth]{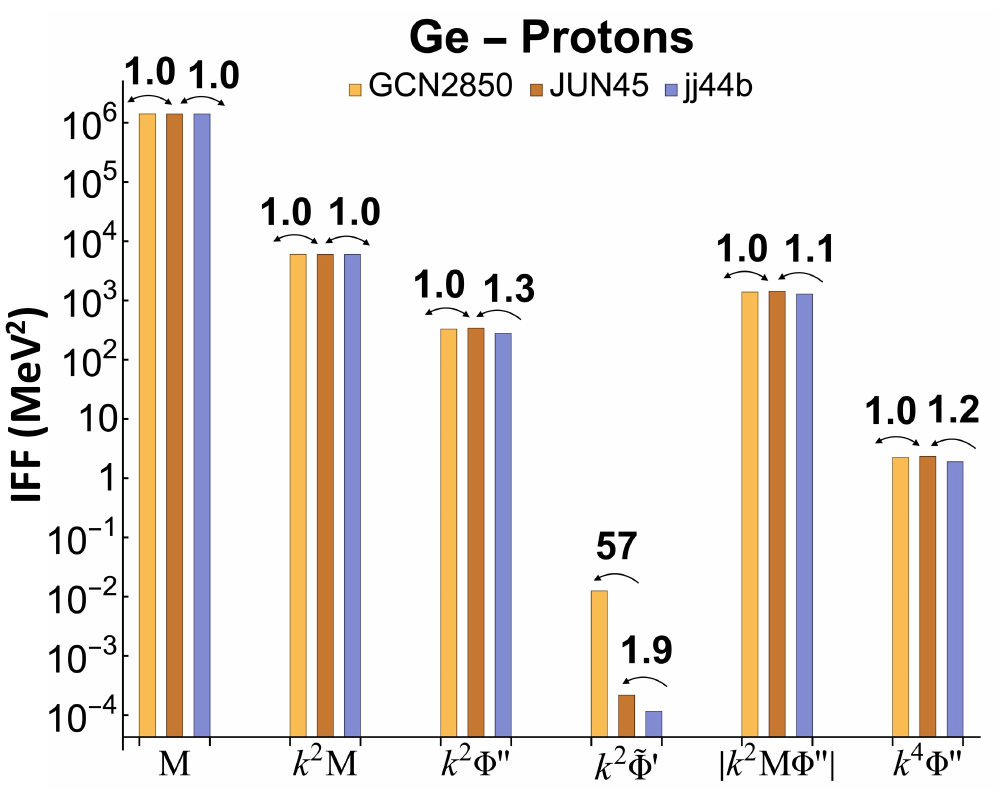}~
    \includegraphics[width=0.355\linewidth]{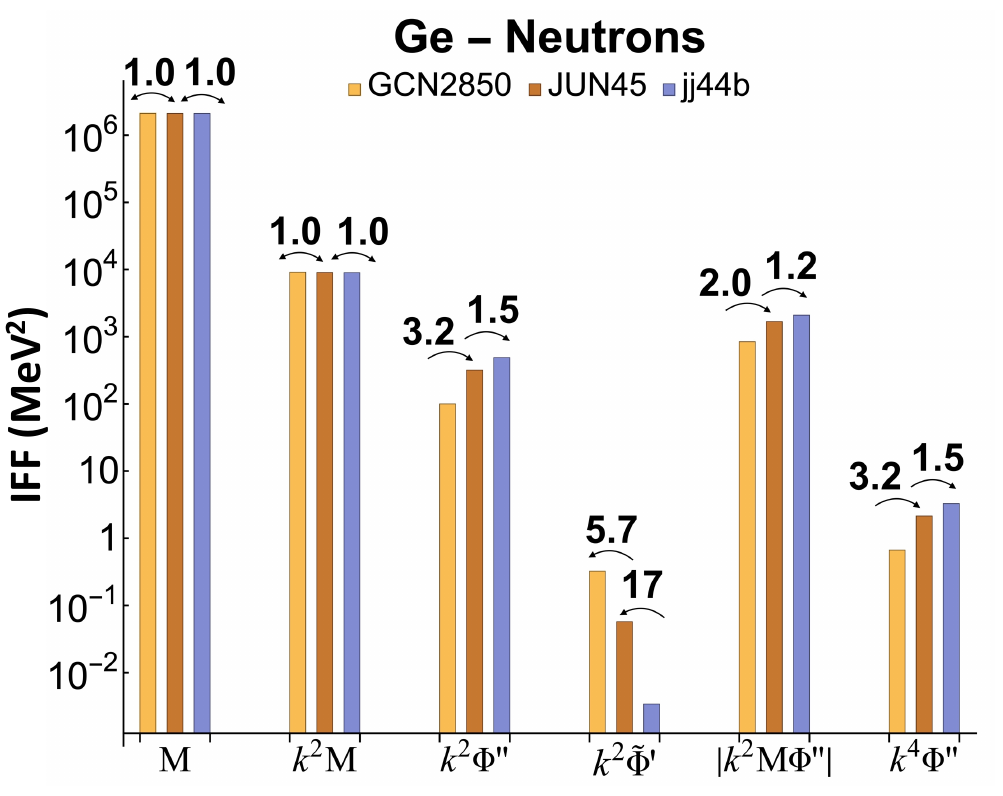}\\
    \includegraphics[width=0.355\linewidth]{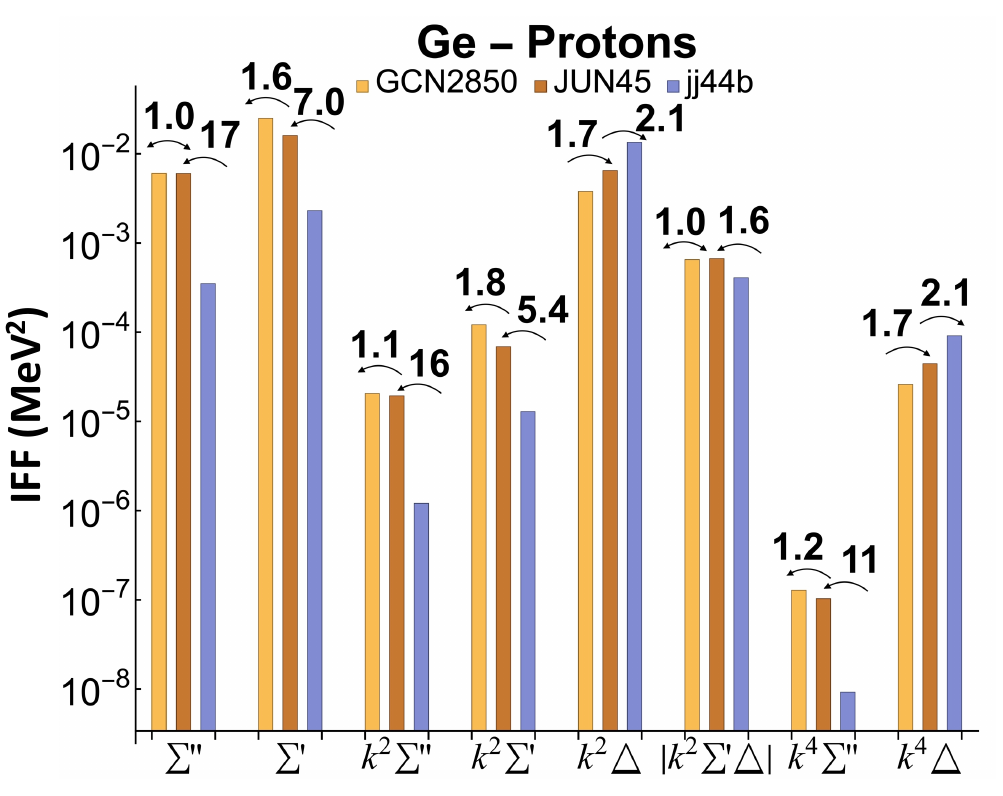}~
    \includegraphics[width=0.355\linewidth]{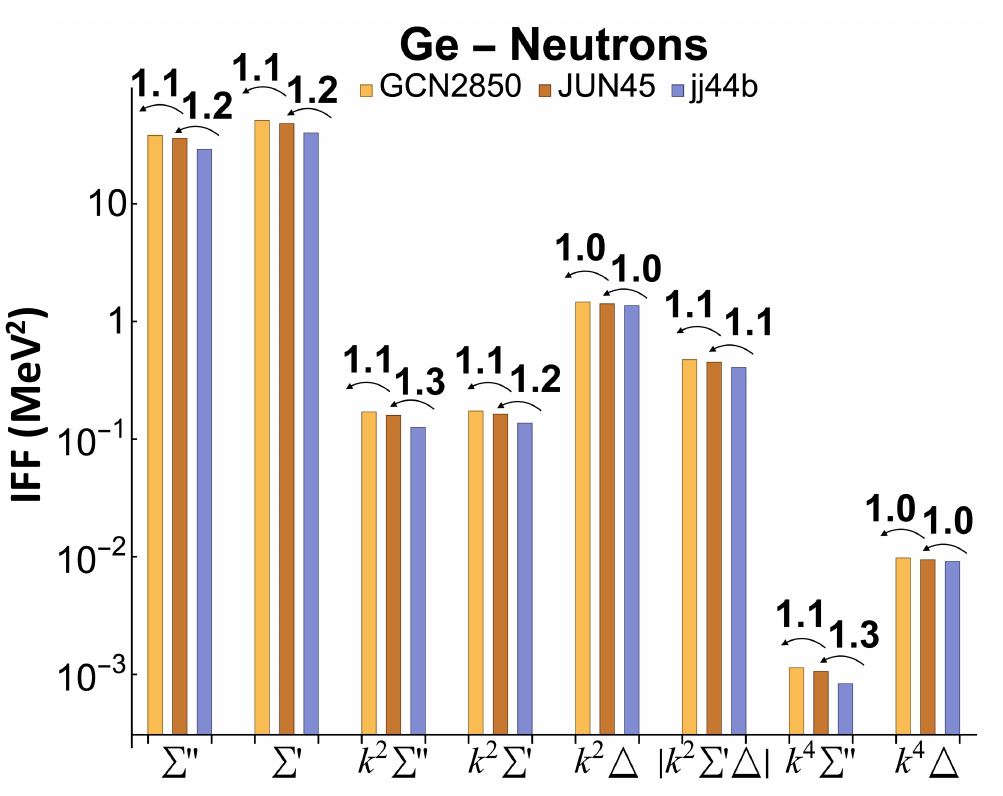}
    \captionsetup{justification=Justified}
    \caption{Germanium proton (left) and neutron (right) IFF values for an integral with $q_{\rm max}= 100$ MeV. Top panel shows values related to the nuclear responses $M$, $\Phi''$, $\tilde{\Phi}'$, whilst bottom panel shows those related to $\Sigma''$, $\Sigma'$, $\Delta$.}
    \label{fig: germanium 100 MeV IFFs}
\end{figure*}

The proton and neutron $q_{\rm max} = 100$ MeV IFF values for silicon and germanium are presented in Figs.~\ref{fig: silicon 100 MeV IFFs} and \ref{fig: germanium 100 MeV IFFs}, respectively. Values on top of IFF bars indicate factor differences between the shell model calculations, with the arrows indicating direction of multiplication.

\begin{table*}[htpb] 
\centering
\captionsetup{justification=Justified}
\caption{Factor differences between the IFF nuclear uncertainties obtained using $q_{\rm max}=50$ MeV and $q_{\rm max}=100$ MeV.
\label{tab: IFF q dependence}}
\begin{tabular}{|l||c|c||c|c|c|c|}
\hline 
\multicolumn{7}{|c|}{\textbf{$\mathbf{q_{\rm max}=50}$ MeV / $\mathbf{q_{\rm max}=100}$ MeV}}\\
\hline
\textbf{Nucleus}  &  \multicolumn{2}{c||}{\textbf{Si}}  & \multicolumn{4}{c|}{\textbf{Ge}}   \\
\hline
 &  \multicolumn{2}{c||}{\textbf{USDB/USD}} & \multicolumn{2}{c|}{\textbf{GCN2850/JUN45}} & \multicolumn{2}{c|}{\textbf{JUN45/jj44b}}  \\
 \hline
 \textbf{$N$}  &  \hspace{3mm} p \hspace{3mm} & \hspace{3mm} n \hspace{3mm} & \hspace{5mm} p \hspace{5mm} & \hspace{5mm} n \hspace{5mm} & \hspace{5mm} p \hspace{5mm} & \hspace{5mm} n \hspace{5mm} \\
 \hline 
 \hline
 $M$ & 1.0 & 1.0  & 1.0 & 1.0 & 1.0 & 1.0 \\
 $\Sigma''$  & 1.4 & 1.0 & 1.0 & 1.0 & 0.8& 1.0 \\
 $\Sigma'$  & 0.8 & 1.0 & 0.8 & 1.0 & 1.4 & 1.0 \\
 \hline
 $q^2 M$ & 1.0 & 1.0  & 1.0 & 1.0 & 1.0 & 1.0 \\
 $q^2 \Sigma''$ & 1.7 & 1.0  & 0.9 & 1.0 & 1.0 & 0.9 \\
 $q^2 \Sigma'$ & 0.7 &  1.0 & 0.7 & 1.0 & 1.8 & 1.0 \\
 $q^2 \Phi''$ & 1.0 & 1.0 & 1.0  & 1.0 & 0.9 & 1.0 \\
 $q^2 \Delta$ & 1.0 & 1.0 & 1.0 & 1.0 & 1.1 & 1.0 \\
 $q^2 \tilde{\Phi}'$ & -- & -- & 0.7 & 1.0 & 0.8 & 0.8\\
 \hline
 $q^4 \Sigma''$ & 1.8 & 1.0  & 0.8 & 1.0 & 1.6 & 0.9 \\
 $q^4 \Phi''$ & 1.0 & 1.0 &  1.0 & 1.0 & 1.0 & 1.0 \\
 $q^4 \Delta$ & 1.0 & 1.0 & 1.0 & 1.0 & 1.1 & 1.0 \\
 \hline
 \end{tabular}
 \end{table*}

Table~\ref{tab: IFF q dependence} showcases the relative difference in the IFF nuclear uncertainties between the $q_{\rm max}=50$ MeV and $q_{\rm max}=100$ MeV calculations, for both silicon and germanium. Changing the upper limit of the IFF integral does not seem to have an impact on the shell model interaction nuclear uncertainties for the neutron nuclear channels of silicon (USDB/USD) and germanium (GCN2850/JUN45). Most of the germanium JUN45/jj44b neutron channels also display no difference between the two IFF integral values, with the exception of $q^2 \Sigma''$ and $q^4 \Sigma''$ which display a difference of $\sim 10 \%$, and $q^2 \tilde{\Phi'}$ which displays a difference of $\sim 25\%$. However, overall the nuclear uncertainties in the neutron channels for silicon and germanium seem to be less sensitive to the $q_{\rm max}$-dependence of the IFF integral. On the other hand, the proton channels of both nuclei are sensitive to this dependence. This is present for the SD channels for silicon, with the largest differences displayed by $q^2 \Sigma''$ ($\sim 70\%$) and $q^4 \Sigma''$ ($\sim 80\%$). The proton germanium SD channels also display a $q_{\rm max}$-dependence in the nuclear uncertainty, with a maximum difference of $\sim 40\%$ between GCN2850 and JUN45, and a maximum difference of $\sim 80\%$ between JUN45 and jj44b. The proton tensor channel $\tilde{\Phi}'$ also displays a difference of $\sim 25-45\%$ (GCN2850/JUN45, JUN45/jj44b), with the $l$-dependent $q^2 \Delta_p$, $q^4 \Delta_p$ channels displaying a $\sim 10\%$ difference (JUN45/jj44b).

Depending on whether the proton or neutron particle coefficients are dominant, this $q_{\rm max}$-dependence of the IFF nuclear uncertainties may not be important for the DM cross section nuclear uncertainty analysis. 

\end{document}